\newcommand{\ndy}{(Nd$_x$Y$_{1-x}$)$_{2}$BaNiO$_{5}$}
\newcommand{\nd}{Nd$_{2}$BaNiO$_{5}$}
\newcommand{\rr}{$R_{2}$BaNiO$_{5}$}
\newcommand{\nii}{Ni$^{2+}$}
\newcommand{\ndd}{Nd$^{3+}$}
\newcommand{\rion}{$R^{3+}$}

\documentstyle[prb,aps,preprint]{revtex}
\begin{document}
\draft

\title{Magnetic ordering, spin waves, and Haldane gap excitations in \protect\ndy\
linear-chain mixed-spin antiferromagnets.}

\author{T. Yokoo\footnote{permanent address : Department of Physics,
Aoyama-Gakuin University, 6-16-1, Chitosedai, Setagaya-ku, Tokyo 157
Japan} S. Raymond\footnote{permanent address : DRFMC/SPSMS/MDN Centre
d'Etudes Nucleaires, 17 rue des Martyrs, 38054 Grenoble Cedex, France.},
A. Zheludev, and S. Maslov}

\address{Physics Department, Brookhaven National Laboratory, Upton,
New York 11973 USA}

\author{E. Ressouche}
\address{DRFMC/SPSMS/MDN Centre
d'Etudes Nucleaires, 17 rue des Martyrs, 38054 Grenoble Cedex, France.}

\author{I. Zaliznyak\footnote{Department of Physics and Astronomy,
Johns Hopkins University, MD 21218 USA,\\ and P.~L.~Kapitza Institute for
Physical Problems, Moscow, Russia.}, R. Erwin}
\address{NIST
Center for Neutron Research, National Institute of Standards and
Technology, MD 20899}

\author{M. Nakamura and J. Akimitsu}

\address{Department of Physics, Aoyama-Gakuin University, 6-16-1,
Chitosedai, Setagaya-ku, Tokyo 157 Japan}

\date{\today}

\maketitle

\begin{abstract}
Linear-chain nickelates with the composition \ndy\ ($x=1$, $x=0.75$,
$x=0.5$, and $x=0.25$) are studied in a series of neutron scattering
experiments. Powder diffraction is used to determine the temperature
dependence of the magnetic structure in all four systems. Single-crystal
inelastic neutron scattering is employed to investigate the temperature
dependence of the Haldane-gap excitations and low-energy spin waves
in the $x=1$ compound \nd. The results of these experiments
are discussed in  the context of the ``Haldane chain in a staggered
field'' model for $R_2$BaNiO$_5$ systems, and quantitative agreement
with theory is obtained.
\end{abstract}

\narrowtext

\section{introduction}

The quantum-disordered ground state and the famous Haldane energy gap in
the magnetic excitation spectrum\cite{Haldane83} have kept 1-dimensional
(1-D) Heisenberg antiferromagnets (AF) at the center of attention of
condensed matter physicists for the last 15 years. Among the more recent
developments are studies of  such systems in external magnetic
fields.\cite{Kakurai91,Regnault92,Kobayashi94,Regnault97} It was found
that in sufficiently strong fields one of the three Haldane-gap modes
undergoes a complete softening.  The result is a transition to a new
phase with long-range antiferromagnetic correlations.\cite{Sakai93,Sorensen93}
The effect of a {\it staggered} field ${\bf H}_{\pi}$, to which a
Haldane spin chain is most susceptible,  is expected to be no less
dramatic. Unfortunately, this problem has been given much less attention
in literature, simply because such conditions are almost impossible to
realize in an experiment. An {\it arbitrarily small} staggered field
will induce a non-zero staggered magnetization in the Haldane chain. An
intriguing question is how this induced order will affect the triplet
gap excitations. Until recently the only 1-D integer-spin AF  whose
behavior was described in terms of a ``staggered field'' is NENP, one of
the best-known Haldane-gap  materials
[Refs.~\onlinecite{Regnault94,Ma92} and references therein]. Due to the
existence of two non-equivalent Ni sites in each chain, a weak effective
staggered field is induced in NENP by a {\it uniform} external
field.\cite{Chiba91,Mitra94} While the main effect on the spin dynamics
is that of the uniform field, the staggered component produces a small
yet observable effect.\cite{Mitra94,Sakai94}

The recent discovery of coexistence of Haldane gap excitations  and
long-range magnetic order in $R_{2}$BaNiO$_{5}$ ($R$=magnetic rare
earth) compounds\cite{Zheludev96PBANO,Zheludev96NBANO,Yokoo97NDY}
presents a unique opportunity to investigate the effect of a staggered
field experimentally without having it obscured by the response of the
system to a  uniform component. According to our current understanding
of the $R_{2}$BaNiO$_{5}$'s, long-range magnetic ordering does not
destroy the quantum-disordered ground state of individual $S=1$ Ni-chains
in these quasi-1-D systems. It is therefore essential that the Ni$^{2+}$
ions carry an {\it integer}, as opposed to half-integer spin, and
qualitatively different behavior is expected for isostructural
half-integer spin materials like
Nd$_{2}$BaCoO$_{5}$.\cite{Hernandez97,Raymond96} The non-zero staggered
magnetization that appears on the intrinsically disordered $S=1$
Ni-chains below the N\'{e}el temperature $T_{N}$ is viewed as being {\it
induced} by an effective staggered exchange field generated by the
ordered $R$-sublattice.\cite{Zheludev96PBANO,MZ} This physical picture
will be referred to as the ``Haldane Chain in a Staggered Field''
(HC/SF) model for $R_{2}$BaNiO$_{5}$ compounds.

In the  HC/SF model, even in the magnetically ordered state, much of the
dynamic spin correlations are contained in  Haldane gap modes
propagating on the Ni chains. Below $T_{N}$ these excitations coexist
with low-energy order-parameter excitations, i.e., conventional spin
waves, that involve correlated fluctuations of both Ni and rare earth
moments. The first experimental confirmation of this was obtained in
inelastic neutron scattering experiments on
Pr$_{2}$BaNiO$_{5}$,\cite{Zheludev96PBANO} where Haldane gap  modes were
clearly seen {\it both above and below the N\'{e}el temperature}. At all
temperatures these excitations have a purely  1-dimensional dynamic
structure factor, and show no sign of softening at the 3-D AF
zone-center at $T=T_{N}$. Pr$_{2}$BaNiO$_{5}$ samples are extremely
difficult to prepare, and their magnetic properties are not always
reproducible. For example, while magnetic ordering at $T_{N}=24$~K has
been clearly observed in single crystals,\cite{Zheludev96PBANO} for
reasons yet unclear, no magnetic transition  was observed in
powders.\cite{Garcia95} It turned out that Nd-based linear-chain
nickelates with the general formula (Nd$_x$Y$_{1-x}$)$_{2}$BaNiO$_{5}$ are
more practical as model systems for neutron studies. On one side of this
series ($x=0$) is Y$_{2}$BaNiO$_{5}$, a well-studied  Haldane-gap $S=1$
antiferromagnet with a spin gap $\Delta
\approx 9$~meV.\cite{Darriet93,DiTusa94,Yokoo95,Sakaguchi96,Xu96} Y$_{2}$BaNiO$_{5}$
remains magnetically disordered even at low temperatures. At the other
end we have Nd$_{2}$BaNiO$_{5}$ ($x=1$) that, thanks to the presence of
magnetic Nd$^{3+}$ ions, orders antiferromagnetically at $T_{N}=48$~K.
As in the case of Pr$_{2}$BaNiO$_{5}$, Haldane-gap modes in
Nd$_{2}$BaNiO$_{5}$ were found to survive  in the magnetically ordered
phase.\cite{Zheludev96NBANO} Similar behaviour was seen in $x=0.75$,
$x=0.5$ and $x=0.25$ species, where the ordering temperatures are 39~K,
30~K, and 19~K, respectively.\cite{Yokoo97NDY} For all Nd concentrations
at $T>T_{N}$ the Haldane gap modes are virtually indistinguishable from
those seen in Y$_{2}$BaNiO$_{5}$. As the temperature is decreased
through $T_{N}$ however, the gap energy $\Delta$ {\it increases} roughly
linearly with $T_{N}-T$. This remarkable behaviour is in agreement with
theoretical predictions based on HC/SF model.\cite{MZ,Zheludev97INV}

To date, all inelastic measurements on (Nd$_x$Y$_{1-x}$)$_{2}$BaNiO$_{5}$
materials were performed on powder samples. The existing magnetic
structure (diffraction) data are not accurate or not complete enough to
be discussed in relation to our model. Indeed, previous single-crystal
experiments on the Nd-system\cite{Sachan94,Zheludev96NBANOX} were
analyzed assuming that the Nd and Ni moments are collinear at all
temperatures below $T_{N}$. It was later shown that the actual structure
is canted,\cite{Matres97} but no detailed temperature dependencies
derived from the canted model were reported. For the mixed Nd-Y systems
only very preliminary results obtained by analyzing only three magnetic
reflections are available.\cite{Yokoo97NDY}

In this work we continue our studies of (Nd$_x$Y$_{1-x}$)$_{2}$BaNiO$_{5}$
compounds using both  elastic and inelastic neutron scattering. Powder
neutron  diffraction is used to investigate the temperature evolution of
the magnetic structure in four (Nd$_x$Y$_{1-x}$)$_{2}$BaNiO$_{5}$  species
with $x=1$, $x=0.75$, $x=0.5$ and $x=0.25$, respectively. Inelastic
neutron scattering is then employed to study the temperature dependence
of the Haldane-gap excitations in the Ni-chains as well as the
low-energy  order-parameter fluctuations (spin waves) in single-crystal
Nd$_{2}$BaNiO$_{5}$ samples. The connection between static and dynamic
properties is discussed in the framework of the HC/SF model. Some of the results
presented below were briefly discussed in Ref.~\onlinecite{mf}.

\section{Experimental procedures}
The crystal structure of $R_{2}$BaNiO$_{5}$  compounds is discussed in
great detail elsewhere.\cite{Garcia93s} All
(Nd$_x$Y$_{1-x}$)$_{2}$BaNiO$_{5}$ systems crystallize in an
orthorhombic unit cell, space group {\it Immm}, $a \approx
3.8$~\AA,~$b\approx 5.8$~\AA,~and $c\approx 11.7$~\AA.\cite{Yokoo97NDY}
The $S=1$ Ni chains are formed by orthorhombically distorted NiO$_{6}$
octahedra that are lined up in the $[100]$ crystallographic direction,
sharing their apical oxygen atoms. The $R$ sites (occupied at random by
Y  or Nd) are positioned in between these chains. The site symmetry for
the $R^{3+}$ is low and the electronic ground state for Nd$^{3+}$ is a
Kramers doublet.

In powder diffraction experiments on the  $x=0.75$, $x=0.5$ and $x=0.25$
compounds  we used the same samples (roughly 15~g each) as in previous
inelastic studies.\cite{Yokoo97NDY} A new Nd$_{2}$BaNiO$_{5}$ ($x=1$)
powder sample of comparable mass was prepared particularly for the
present study using the solid state reaction method. Powder neutron
diffraction experiments were carried out at the high flux reactor of the
Institut Laue-Langevin (Grenoble), on the 400 cells position sensitive
detector diffractometer D1B for the $x=0.75$, $x=0.5$ and $x=0.25$
compounds and on the 1600 cells diffractometer D20 for the $x=1$
system.  The patterns were recorded in the temperature range 2 - 50 K,
using a liquid helium cryostat.  The wavelengths $\lambda = 2.526$
\AA~on D1B and $\lambda = 2.400$ \AA~on D20 were provided
in both cases by focusing pyrolitic graphite monochromators.  The
samples were enclosed in cylindrical vanadium containers. The analysis
of the powder patterns was performed by Rietveld profile refinement
using the software FULLPROF.\cite{Rodriguez93} A pseudo-Voigt function
was chosen to generate the line shape of the diffraction peaks.  The
scattering lengths were taken from Ref.~\onlinecite{Koester91} and the
magnetic form factors of Nd and Ni from Ref.~\onlinecite{Brown92b}

Single-crystal samples for inelastic neutron scattering experiments were
grown in oxygen atmosphere  using the floating-zone technique.  Most of
the  thermal neutron inelastic measurements were done on a cylindrical
single crystal 25~mm long and 4~mm in diameter (sample A) with  mosaic
about $0.3^{\circ}$. Only  near the end of this series of experiments
did a much larger sample (sample B) become available. Sample B
consists of three single crystals, each roughly 6~mm in diameter and 50 mm long
with an $\approx 0.8^{\circ}$ mosaic. The three crystals were aligned
together on a specially designed aluminum sample holder. The resulting
``supersample'' had a symmetric mosaic of around $1.2^{\circ}$.
Sample B was used primarily in cold-neutron inelastic scattering
experiments.

Inelastic thermal neutron scattering  measurements were carried out at
the Center for Neutron Research at the National Institute of Standards
and Technology (NIST) on the BT-4 and BT-2 triple-axis spectrometers.
The samples were in all cases mounted with the $(h,0,l)$
reciprocal-space plane parallel to the scattering plane of the
spectrometer. A neutron beam of $14.7$~meV fixed final energy was used
with $60'-40'-40'-120'$ collimations and a pyrolitic graphite (PG)
filter positioned after the sample. PG (002) reflections were used for
monochromator and analyzer (setup I). The measurements were done with
energy transfers up to 25~meV, the calculated resolution at a typical
15~meV energy transfer being 2.5~meV FWHM. The sample environment was a
standard displex refrigerator, and measurements were done in the
temperature range 25--60~K. As we have found, in all samples the axis of
the cylindrical crystal roughly coincides with $[100]$ crystallographic
direction. In the experiment the sample cylinder was therefore
positioned horizontally in the scattering plane, which resulted in
substantial absorbtion effects. The effective transmission coefficient
as a function of sample orientation was determined by measuring the
intensity of incoherent scattering. The measured absorption correction
was almost constant within each constant-$Q$ scan performed, but
varied by as much as a factor of two between different Brillouin zones.

Inelastic cold neutron scattering experiments were done on the SPINS
3-axis spectrometer at NIST.  Several configurations were used. In setup
II we utilized $30'-80'-80'-{\rm open}$ collimations with a $3.7$~meV
fixed final energy neutron beam and a BeO filter after the sample. In
this  setup the energy resolution at zero energy transfer is $0.2$~meV
FWHM. $5$~meV fixed-final energy neutrons were used in setup III with
the same collimation setup and a Be-filter positioned after the sample.
A substantial gain in intensity (at the cost of wave vector resolution)
was obtained using $30'-80'-45'({\rm radial})-{\rm open}$ collimations
and $5.1$~meV fixed final energy neutrons, a Be-filter after the sample
and a horizontally-focusing analyzer (setup IV). The energy resolution
in this case was $0.35$~meV FWHM at zero energy transfer. The sample
environment was a displex refrigerator and the temperature was
controlled in the range 10--50~K.

\section{Experimental results}
\subsection{Magnetic structure}
For all samples studied the diffraction patterns in the paramagnetic
phase were found to be totally consistent with the crystal structure
reported in Ref.~\onlinecite{Garcia93s}. A small amount of non magnetic
impurities was detected in the $x=0.75$, $x=0.5$ and $x=0.25$ samples,
whereas the $x=1$ one turned out to be single phase. The small extra
peaks due to this impurity were excluded from the refinements.

As previously observed,\cite{Yokoo97NDY} magnetic ordering in all
samples manifests itself in the appearance of new Bragg reflections at
half-integer positions, the magnetic propagation vector being $[1/2~0~1/2]$.
In the $x=1$ sample, thanks to the relatively strong magnetic
signal, a simultaneous refinement of nuclear and magnetic structures could be performed. 
A typical powder scan for this system is shown in Fig.\ref{difdata}(a). 
For the other three compounds, where the magnetic signal was weaker, a different approach
was used. For powder scans collected at $T<T_{N}$ the magnetic contribution to scattering was isolated
by subtracting the nuclear background measured just above the ordering temperature.
Several typical data sets obtained in this fashion are shown in
Fig.\ref{difdata}(b-d).

In the $x=0.25$ sample, in addition to the above mentioned magnetic
peaks at half-integer positions, extra intensity below $T_{N}$ was
also observed at the positions of nuclear peaks
[Fig.\ref{difdata}(d)].  One possible explanation for this is the
presence of an extra magnetic component in the structure of this
compound.  This however appears not to be the case.  We were unable to
reproduce the observed pattern with any reasonable spin
model.  In addition, the intensities of these extra reflections seem
to {\it increase} with increasing scattering angle, strongly suggesting that they
are of nuclear, rather than magnetic origin.  In our analysis we have
assumed that theses extra peaks represent a lattice distortion that is
induced by magnetic ordering.  In the determination of the magnetic
structure of this sample these reflections were therefore
ignored.

According to the analysis of the crystallographic space group of R$_{2}$BaNiO$_{5}$
compounds ($Immm$), only two types of magnetic structure with
propagation vector $[1/2~0~1/2]$ and non-zero magnetic moment on both Ni and $R$ sites
are possible.\cite{Garcia93}  In the structure realized in Er$_{2}$BaNiO$_{5}$ all spins are
confined to the $(a,b)$ crystallographic plane.  As shown by several
previous studies,\cite{Sachan94,Matres97} it is the other alternative
that is realized in Nd$_{2}$BaNiO$_{5}$: both the Ni$^{2+}$ and
Nd$^{3+}$ moments are in the $(a,c)$ plane of the crystal
(Fig.~\ref{cartoon}).  We used this model to fit the diffraction spectra
measured at each temperature for each sample.  The adjustable parameters
were the magnetic moments $m^{({\rm Ni})}$ and $m^{({\rm Nd})}$ of the
Ni$^{2+}$ and Nd$^{3+}$ ions, respectively, as well as the angles
$\phi^{({\rm Ni})}$ and $\phi^{({\rm Nd})}$ these moments form with the
$c$ axis of the crystal.  Good fits to the experimental powder profiles,
examples of which are shown in solid lines in Fig.~\ref{difdata}, were
obtained in all cases.  The refined values of parameters are plotted
against temperature in Fig.~\ref{MvsT} and Fig.~\ref{PvsT}.  As observed
in previous studies,\cite{Matres97} near the N\'{e}el temperature the
magnetizations of both Ni and Nd sublattices undergo a slight
reorientation with decreasing temperature.  Very quickly though this
reorientation slows down and in all samples the angles $\phi^{({\rm
Ni})}$ and $\phi^{({\rm Nd})}$ remain at roughly $-35^{\circ}$ and
$0^{\circ}$, respectively, in most of the temperature range.

\subsection{Ni-chain gap excitations}
The temperature dependencies of energies and intensities of the
Haldane-gap excitations were studied by inelastic thermal neutron
scattering. One of our objectives was to use the intrinsic polarization
dependence of the neutron scattering cross section to distinguish
between  the three individual modes, expected to be present in the
Haldane multiplet. For this purpose  the data were collected in
constant-$Q$ scans at the 1-D antiferromagnetic zone-centers ${\bf
Q}=(1.5,0,0)$ and ${\bf Q}=(0.5,0,4.2)$. These wave vectors correspond
to equal momentum transfers, but are pointing along and almost
perpendicular to the Ni-chain axis $a$, respectively. At $(1.5,0,0)$
only the fluctuations of $y$ and $z$ spin components are seen (the axes
$x$, $y$ and $z$ are chosen along the $[100]$, $[010]$ and $[001]$
directions, respectively), while  at $(0.5,0,4.2)$ it is mostly the
fluctuations of $x$ and $y$ spin components that contribute to the
scattering intensity.

The most serious obstacle  in previous powder measurements was the
presence of two crystal field (CF) excitations associated with Nd$^{3+}$
that appear at about $18$~meV and $24$~meV,
respectively.\cite{Zheludev96NBANO} Unlike the highly dispersive Haldane
excitations, the dispersionless CF modes do not suffer an intensity
penalty upon powder averaging the corresponding dynamic structure
factors. The Haldane modes that appear at $\approx 10$~meV energy
transfer at $T=50$~K and move towards higher energies at lower
temperature are thus very difficult to separate from the intense CF
background. In a single crystal this problem may be overcome. First, the
intensity of Haldane modes relative to those of CF excitations is much
larger in this case. Second, in a single crystal the  background can be
directly measured at scattering vectors slightly off the 1-D
zone-center, where the Haldane excitations move out of the scan range
thanks to their steep dispersion. In our case the background was
measured at ${\bf Q}=(1.4,0,0)$ and ${\bf Q}=(0.4,0,4.2)$. Since it was
previously found that the CF modes are slightly temperature-dependent,
the background  measurements were performed separately at each
temperature. Figure~\ref{exdata} shows some typical data collected in
sample A at $T=50$~K (open circles) together with the corresponding
background scans (solid circles). The result of point-by-point
subtraction is shown in Fig.~\ref{exsub}. The  well-defined peak at
$\approx 11$~meV energy transfer, very similar to that previously seen
in Y$_2$BaNiO$_5$\cite{Darriet93,DiTusa94,Yokoo95,Sakaguchi96,Xu96} and
Pr$_2$BaNiO$_5$,\cite{Zheludev96PBANO} is attributed to the Haldane
excitations in the Ni chains.

One  problem that had to be dealt with was the presence of a spurious
peak at roughly $18$~meV energy transfer (see Fig.~\ref{exdata}). We
have identified this spurion as being caused by higher-order scattering
($2k_{i}$) in the monochromator, incoherent and/or Bragg scattering in
the sample, and higher-order ($3k_{f}$) scattering in the analyzer. In
the analysis  described below  the ``dangerous'' energy range has been
excluded from all data sets.

Ideally, for a complete  determination of the mode polarization, at each
temperature one would analyze the inelastic scans collected at both wave
vectors simultaneously, using a model cross section that would include
three modes with separate gap values and polarization factors.
In the present study however,  as the structure of the observed peak is
totally smeared by the  broad instrumental resolution, using a
three-component model cross section results in
an over-parameterized problem. Instead we analyzed all the measured
inelastic scans separately using a single-mode cross-section, as was
previously done for Pr$_2$BaNiO$_5$.\cite{Zheludev96PBANO} The model
dynamic structure factor was written in the ``double-Lorentzian'' form:
\begin{eqnarray}
S(\tilde{q},\omega)=\frac{S_{0}\xi/\Gamma}{1+\tilde{q}^2\xi^2+(\omega-\omega_{{\bf
q}})^2/\Gamma^2}\label{sqw}\\ (\hbar
\omega_{{\bf q}})^2=\Delta^{2}+c^{2}\tilde{q}^2.\label{disp}
\end{eqnarray}
In this formula $\tilde{q}$ is the projection of the scattering vector
${\bf q}$ onto the chain axis, and measured relative to the 1-D
antiferromagnetic zone-center; $\Delta$ is the gap energy; $c$ is the
spin wave velocity along the chains; $\Gamma$ is the intrinsic energy
width of the excitations; and $\xi$ is the real-space dynamic spin
correlation length. Since all measurements were performed at
$\tilde{q}=0$, the parameters $\xi$ and $c$ are only needed to properly
take into account focusing effects. In  our analysis $\xi$ and $c$ were
therefore fixed to the values previously measured in Pr$_2$BaNiO$_5$:
$1/\xi=0.08$~\AA$^{-1}$, and $c=200$~meV~\AA. For each scan the cross
section (\ref{sqw}) was folded  with the 4-dimensional spectrometer
resolution function and the parameters $\Delta$, $\Gamma$ and the
prefactor $S_{0}$ were refined to best-fit the data. To further minimize
the number of  independent variables we have averaged the value of
$\Gamma$ determined at different temperatures and used this average as a
fixed parameter in the final fit. This was done separately for ${\bf
Q}=(1.5,0,0)$ and ${\bf Q}=(0.5,0,4.2)$, where the average values were
$2$~meV and $1.5$~meV, respectively. Several typical fits of
Eqs.~(\ref{sqw},\ref{disp}) to our inelastic data are shown in solid
lines Fig.~\ref{exsub2}. The obtained temperature dependencies for
$\Delta$ and $S_{0}$ for both wave vectors are shown in Fig.~\ref{vst}.

\subsection{Low-energy spin waves}
The main purpose of the cold-neutron experiments was to study the
low-energy  order-parameter fluctuations in Nd$_2$BaNiO$_5$.
Usually,
such excitations  resulting from long-range ordering appear as acoustic
spin waves, as previously observed in
Pr$_2$BaNiO$_5$.\cite{Zheludev96PBANO} For Nd$_2$BaNiO$_5$ most of the
measurements were done in the vicinity of the $(0.5,0,1.5)$ magnetic
Bragg reflection, one of the strongest magnetic peaks in the $(a,c)$
plane, with roughly 10000 counts per second at $T=10$~K using setup II. To
search for the acoustic spin waves we used configurations I, II or III
to perform constant-$Q$ scans at wave vectors $(0.5,0,1.5)$,
$(0.75,0,1.5)$, $(1,0,1.5)$, and $(1.5,0,1.25)$ in the range 0--8~meV
(Fig.~\ref{nowave}), as well as constant-$E$ scans at an energy transfer
$\hbar
\omega=0.3$~meV along $(0.5+\zeta,0,1.5)$ and $(0.5,0,1.5+\zeta)$ ($-0.25\lesssim
\zeta\lesssim 0.25$). At $T=10$~K absolutely no inelastic signal was found
at energy transfers  below $\approx 4$~meV. At $\hbar \omega= 4$~meV we
observe the rather intense resolution-limited inelastic feature
previously seen in powder experiments.\cite{Zheludev96NBANO} The $4$~meV
excitation  is totally dispersionless along both the $a$ and  $c$ axes
of the crystal. This can be deduced from Fig.~\ref{spinsdata1}, that
shows some typical constant-$Q$ scans collected at several wave vectors
at $T=10$~K.  Scanning through the 4~meV  peak in different Brillouin zones
led us to the conclusion that the mode is a
single-ion excitation with no unit-cell structure-factor and no apparent
polarization-dependence of intensity. The accessible range of momentum
transfers was not sufficient for an accurate study of the form-factor
dependence of the intensity.

What makes the 4~meV mode rather interesting is its temperature
dependence. As was previously established in inelastic neutron
scattering experiments on powder samples,\cite{Zheludev96NBANO} this
excitation is visible only at $T<T_{N}$. As the N\'{e}el temperature is
approached from below the excitation energy decreases, the width
increases dramatically (Fig.\ref{spinsdata2}), and at $T=T_{N}$ the mode
appears to merge with quasielatic scattering at $\hbar \omega=0$.  We
have utilized setups I and IV to measure the temperature dependence of
the excitation energy for temperatures up to $T=35$~K, where the
inelastic peak is broad, yet still easily identified. Constant-$Q$ scans
were analyzed with a  damped-oscillator dynamic structure factor:
\begin{equation}
S(\omega)=\frac{I_{0}\gamma\left[n(\omega)+1\right]}{\omega_{0}\pi}\left[
\frac{1}{(\omega-\omega_{0})^2+\gamma^2}-\frac{1}{(\omega+\omega_{0})^2+\gamma^2}\right],\label{do}
\end{equation}
that was convoluted with  the gaussian energy-resolution of the
spectrometer (solid lines in Fig.\ref{spinsdata2}). In Eq.~\ref{do}
$n(\omega)$ is the Bose factor and $I_{0}$ is the structure factor of
the excitation. For ${\bf Q}=(0.5,0,1.5)$ the experimental
$T$-dependencies of the excitation energy $\hbar\omega_{0}$ and
relaxation rate $\gamma$ are shown in Fig.~\ref{spinsresults}. The same
behavior was also seen at ${\bf Q}=(0.75,0,1.5)$ and ${\bf
Q}=(1,0,1.5)$.

\section{Discussion}
\subsection{Magnetic ordering}
In previous resonant magnetic X-ray diffraction
studies\cite{Zheludev96NBANOX} it was found that the order-parameter
critical exponent $\beta$ in
\nd\ is indistinguishable from the mean-field (MF) value
$\beta=0.5$. In fact, for the $x$=1 system, the standard MF theory can
accurately describe the entire temperature dependence of both Ni and Nd
ordered moments.\cite{Zheludev96NBANOX} In this model {\it both}
Ni$^{2+}$ and Nd$^{3+}$ ions are treated as isolated spins in an
effective exchange field, and their bare magnetization curves are given
by appropriate Brillouin functions. The reduced moment on the Ni-sites
(the expected value is 2~$\mu_{{\rm B}}$) is attributed to orbital
effects (reduced $g$-factor). Despite that reasonably good fits to the experimental temperature
dependencies can be obtained by applying this simple approach to each
\nd\ compound separately, the refined Ni-Nd and Ni-Ni coupling
parameters, as well as the $g$-factor for Ni$^{2+}$ depend greatly upon
sample composition. Such a strong variation of the Ni-Ni exchange
constant is difficult to explain by the slight composition dependence of
the lattice parameters and hence the Ni-Ni exchange
integral.\cite{Yokoo97NDY} Indeed, the Haldane gap energies in the
paramagnetic phase are directly proportional to the Ni-Ni exchange
constant, and were previously found to be independent of Nd
content.\cite{Yokoo97NDY} If we are to describe the measured temperature
dependencies of sublattice magnetizations in
(Nd$_x$Y$_{1-x}$)$_{2}$BaNiO$_{5}$, treating magnetic interactions at
the MF level, we have to look for a different approach.

In-chain Ni-Ni AF exchange coupling is
by far the strongest magnetic interaction in \rr\ compounds ($J_{{\rm Ni-Ni}}\approx 300$~K). At
$T\lesssim\Delta$, which, incidentally is the
temperature range where long-range magnetic ordering occurs,
the susceptibility of the Ni subsystem is therefore dominated by well-established
in-chain correlations. In this situation it is {\it not
appropriate} to model the magnetic transition by treating the
\nii\ ions as isolated moments. Instead, the MF theory should treat
entire Ni-chains as inseparable entities, characterized by their bare
staggered magnetization function.
3-dimensional long-range ordering is driven by the next important
magnetic interaction, namely  Ni-$R$ exchange coupling that establishes links between individual chains and
is of the order of tens of Kelvin. Direct $R$-$R$ interactions are
expected to be very weak (of the order of 1~K or less) and can
for this reason be disregarded.\cite{Garcia95}

Garcia-Matres et al.\cite{Garcia95} were probably
the first who analyzed their magnetic susceptibility data treating only
the Ni-R interactions at the MF level. The experimentally measured
temperature dependence of the Ni-order parameter was built into this
model as an empirical function. The HC/SF model provides a physical
justification for this approach:\cite{mf} the bare magnetization curve for the Ni-subsystem is to be taken
in the form of the {\it staggered magnetization function for an isolated
quantum-disordered Haldane chain}. This ``semi-quantum'' model
differs qualitatively from the standard MF model for classical magnets.
In the latter, at $T\rightarrow 0$ all sublattices become fully
saturated as the bare single-ion susceptibility diverges as $1/T$. In
contrast, the bare staggered susceptibility $\chi_{\pi}$ of a Haldane
spin chain remains {\it finite} at $T=0$, and the Ni-sublattice need not
be fully saturated. The ``semi-quantum'' model can thus account for the
reduced ordered moment on the Ni-sites without assuming an improbably
small $g$-factor for \nii. Moreover, it will predict qualitatively
different temperature dependencies, since the staggered susceptibility
of a Haldane spin chain, unlike the single-ion susceptibility, is
expected to be almost $T$-independent at $T\lesssim \Delta$ ($\Delta
\approx 9$~meV, or $\approx 105$~K in $R_{2}$BaNiO$_{5}$ compounds).

It has to be noted that treating interchain coupling at the
MF or RPA (Random Phase Approximation) level is, in itself, not a new
idea, but a well-established technique\cite{Scalapino75}. In their
pioneering work on CsNiCl$_{3}$, the first Haldane-gap material studied
experimentally, Buyers et al. \cite{BuyersCsNiCl3}, and Affleck
\cite{Affleck89} implemented this approach to explain magnetic ordering
and calculate the spin wave dispersion relations. The main difference
between CsNiCl$_{3}$ and $R_{2}$BaNiO$_{5}$ is that exchange coupling
between individual Haldane chains is {\it direct} in the former system,
and {\it mediated by the rare earth ions} in the latter. For directly
coupled Haldane spin chains, the magnitude of interchain interactions
must exceed some critical value in order for the system to order
magnetically \cite{Affleck89}. In a MF description of CsNiCl$_3$ or any other
system with directly couple quantum chains, the temperature
dependence of the ordered moment is defined by the {\it intrinsic}
temperature dependence of the susceptibility of individual chains. In
contrast, in our case of $R_{2}$BaNiO$_{5}$ compounds, magnetic ordering
is driven by the $1/T$-divergent susceptibility of the rare earth
subsystem. At sufficiently low temperature long-range order will
therefore occur for arbitrary small $R$-Ni interactions, and  in the
case of $T_{N}\lesssim \Delta$, we can use the approximation where the
bare magnetization curve of isolated chains is $T$-independent.

An important results of this work is that in {\it all samples studied}
$m^{({\rm Ni})}$, the induced staggered cmoment on the Ni-chains, appears
to be a universal function of $\tilde{m}^{({\rm
Nd})}\equiv m^{({\rm Nd})} x$. In other words, the Ni magnetization
explicitly depends only on the average moment on the $R$-sites, and not
the actual temperature, as illustrated in the plot of $m^{({\rm Ni})}$
vs. $\tilde{m}^{({\rm Nd})}$ in Fig.~\ref{univ}. To emphasize the
significance of this fact we reiterate that the ordering temperatures in
the materials differ by more than a factor of two.  From Fig.~\ref{univ}
we see  that even with fully ordered  \ndd\ moments the Ni-chains are not
fully polarized and the curve never levels off completely. In the
context of the HC/SF model, apart from the scaling of the absciss,
Fig.~\ref{univ} is nothing else but the {\it staggered magnetization
function} ${\cal M}(H_{\pi})$ for an isolated Haldane chain of Ni$^{2+}$
spins, that in the studied temperature range $T=0-50$~K is expected to
be almost $T$-independent.

The role of disorder in Y-diluted systems  is
not too important, at least, in systems where the Nd
concentration is not too small. The intrinsic dynamic spin correlation length of Haldane
spin chains is rather large,  of the order of  six lattice repeats. As
long as the mean distance between the
\ndd\ ions is smaller than this length scale, the Ni-chains
effectively see a homogeneous staggered moment on the $R$-sites, despite
the ``holes'' that are present wherever \ndd\ is replaced by Y$^{3+}$.
The argument, while clearly valid at $T\approx T_{N}$,  should be
taken with some caution when applied to a system deep in the
ordered phase. In this regime  the Haldane gap energy increases, and, as
a consequence of that the dynamic correlation length in the chains decreases.
Even in this case we can expect the effect of disorder to be averaged out,
thanks to the simple geometrical fact that every Ni$^{2+}$ ion is coordinated to four $R$-sites.

We can now use the experimentally determined staggered magnetization
function for Haldane spin chains in \ndy\ to write down the
self-consistent MF equations for our ``semi-quantum'' theory. For the
Ni-sublattice these equations are:
\begin{eqnarray}
 m^{({\rm Ni})}= g S \mu_{B} {\cal M}(H^{({\rm Ni})})\label{e1}\\
 H^{({\rm Ni})}=2\alpha \tilde{m}^{({\rm Nd})}\label{e2}
\end{eqnarray}
Here $g\mu_{B}$ and $S=1$ are the gyromagnetic ratio and spin of \nii,
respectively; $H^{({\rm Ni})}$ is the effective exchange field that acts
on the Ni-chains and is generated by the $R$-sublattice; and $\alpha$ is
an effective MF coupling constant. The factor 2  in Eq.~\ref{e2}
reflects the  fact that there are two Nd atoms for every Ni atom in the
chemical formula, while the magnetizations $m^{({\rm Ni})}$ and
$m^{({\rm Nd})}$ are normalized per site. For  convenience ${\cal
M}(H^{({\rm Ni})})$ in Eq.~\ref{e1} is approximated with the following
purely empirical function:
\begin{equation}
g S \mu_{B} {\cal M} (2 \alpha \tilde{m}^{({\rm Nd})})= A \arctan (B
\tilde{m}^{({\rm Nd})})\label{fun}.
\end {equation}
The coefficients $A=1.27(5)$ and $B=1.06(13)$ are obtained by fitting
this formula to the bulk of experimental  data in Fig.~\ref{univ} (solid
line). To write down the remaining MF equations for the $R$-sublattice
we shall approximate the bare magnetization curve for the
\ndd\ ions with the  Brillouin function, as is appropriate for a magnetic ion with
a doublet ground state:
\begin{eqnarray}
m^{({\rm Nd})}= m^{({\rm Nd})}_{0} \tanh (\frac {m^{({\rm Nd})}_{0}
H^{({\rm Nd})}}{\kappa T}).\label{e3}\\
 H^{({\rm Nd})}=\alpha m^{({\rm Ni})}\label{e4}.
\end{eqnarray}
In this formula $H^{({\rm Nd})}$ is the  effective exchange field acting
on the rare earth ions, $\kappa$ is Boltzmann's constant and $m^{({\rm
Nd})}_{0}$ is the effective magnetic moment for \ndd.

Ideally, the saturation moment $m_{0}$ should be indepent of temperature
or Nd concentration in the sample. In our particular systems however,
the saturation moment of the Nd$^{3+}$ ions steadily decreases by
roughly 25\% as $x$ changes from 1.0 to 0.25. Eq.~\ref{e3} is therefore
no more than a crude approximation, and a 25\% accuracy is the best we
can expect from our model. Indeed, in using the Brillouin function to
describe the rare earths, we have totally neglected the higher-energy
electronic states of these ions. Not only do they contribute to the
temperature dependence of $R$-magnetization, but also may give rise to a
$T$-independent Van-Vleck contribution to single-ion susceptibility.
Unfortunately, without knowing the electronic structure of \rion\ in
\rr\ in detail, we can not take these effects into account rigorously.
Instead we shall use Eq.~\ref{e3} with separate values  for $m_{0}$ for
each sample, equal to the saturation magnetization of the Nd sublattice.

Equations (\ref{e1}-\ref{e4}) were used to analyze the temperature
dependencies of sublattice magnetizations in three
\ndy\ samples, with $x=1$, $x=0.75$ and $x=0.5$. In the $x=0.25$ system
the transition is of magneto-elastic nature, and a purely magnetic MF
model can not be expected to be applicable. For each sample composition
the only adjustable parameter was the coupling constant $\alpha$. The
results of this analysis are shown in solid lines in Fig.~\ref{MvsT}.
For the three samples we obtain
$\alpha=0.84(2)\cdot10^5$~Oe$\ \mu_{B}^{-1}$,
$\alpha=0.96(2)\cdot10^5$~Oe$\ \mu_{B}^{-1}$, and
$\alpha=1.14(2)\cdot10^5$~Oe$\ \mu_{B}^{-1}$, respectively. As expected,
for all three
\ndy\ systems the refined values of the coupling constant are very similar
despite the substantially different ordering temperatures.  We can use
average value $\alpha=0.98(1)\cdot10^5$~Oe$\ \mu_{B}^{-1}$ to properly
rescale the absciss in Fig.~\ref{univ} and obtain the Ni-order parameter
as a function of effective exchange field, i.e., the actual {\it
staggered magnetization curve for a Haldane spin chain}
(Fig.~\ref{univ}, top axis). Differentiating Eq.~\ref{fun} at $H^{({\rm
Ni})} \rightarrow 0$ we obtain the {\it staggered susceptibility}
$\chi_{\pi}=1.2\cdot 10^{-5}
\mu_{B}$/Oe.

The staggered magnetization curve for an isolated $S=1$ Heisenberg AF
chain has not been calculated to date, and we can not directly compare
our results to any  numerical simulations. The zero-temperature
staggered susceptibility $\chi_{\pi}$, on the other hand, can be deduced
from the numerous theoretical predictions for the dynamic structure
factor at the 1-D AF zone-center $S(\pi,\omega)$, to which it is related
through the Kramers-Kronig relation and the fluctuation-dissipation
theorem. In the single-mode approximation\cite{Arovas88,Muller81,Ma92}
$S(\pi,
\omega)=-\case{4}{3}\case{\langle {\cal H} \rangle}{L \Delta}\delta(\omega-\Delta)$.
Here and $\case{\langle {\cal H} \rangle}{L}$ is the ground state energy
per spin. According to Monte Carlo simulations (see, for example,
Refs.~\onlinecite{Liang90,Meshkov93}), $\case{\langle {\cal H}
\rangle}{L}\approx -1.4 J\approx 3.42 \Delta$. The expression for the staggered
susceptibility is then readily obtained via
$\chi_{\pi}=2\int_{0}^{\infty}d\omega \case{S(\pi,\omega)}{\omega}$ and
is equal to:
\begin{equation}
\chi_{\pi}=9.1 (g \mu_{B})^2/\Delta.\label{estimate}
\end{equation}
In all the $R_{2}$BaNiO$_5$ systems studied so far with inelastic
neutron scattering  the gap energies are  equal to 9~meV to within
experimental error, which gives $\chi_{\pi}\approx 2.3\cdot 10^{-5}
\mu_{B}$/Oe. This value is almost twice as large as our experimental
estimate. While not spectacular, this level of consistency is quite
acceptable, considering all the approximations and simplifications that
had to be made. Most of the uncertainty is related to the large
experimental error in the initial ``low-field'' part of the
magnetization curve in Fig.~12 where small ordered moments were
derived from the measurements in the vicinity of $T_N$, and the error bars are rather large.
Some systematic error is introduced by approximating the measured curve with
the expression (\ref{fun}). We have also totally neglected the
temperature dependence of the magnetization function for the Haldane
spin chains, small as it may be at $\kappa T\lesssim 0.5 \Delta$, and
approximated that for the rare earth ions by a simple Brillouin
function. The actual gyromagnetic ratio for Ni$^{2+}$ that enters
Eq.~\ref{estimate} squared, is not known from independent measurements
either. A half-order-of-magnitude agreement is indeed the most we can
expect. What is important is that our model appears to be
self-consistent and is based on the same concepts as those previously
used to explain the persistence and $T$-dependence of Haldane-gap
excitations in $R_{2}$BaNiO$_5$ materials.

\subsection{The 4 meV mode}
The excitation that we see at 4 meV energy transfer was first observed
in powder experiments,\cite{Zheludev96NBANO} and attributed to a CF
transition in Nd$^{3+}$.  A similar  feature was also seen in recent
AFMR measurements.\cite{Okubo97} As our new single-crystal neutron data
clearly demonstrate, the 4 meV  mode has indeed no dispersion and its
intensity is Brillouin-zone independent, so it indeed is a local, i.e.,
single-ion excitation(s). Despite that, we  suggest
that this mode {\it is} the actual order-parameter excitation (spin wave),
and corresponds to flipping a single Nd$^{3+}$
moment in the effective staggered field projected by the Ni-sublattice.
The energy of such an excitation is given by:
\begin{equation}
\hbar\omega_{0}=2m_{0}H^{({\rm Nd})}=2m_{0}\alpha m^{({\rm Ni})}\label{4mm}.
\end{equation}
The excitation energy should thus  scale as the ordered moment of the
Ni-sublattice. The solid line in Fig.~\ref{spinsresults} is drawn using
Eq.~\ref{4mm}, our experimental data for $m^{({\rm Ni})}(T)$ in \nd, and
$\alpha=0.84\cdot10^{5}$ Oe$\mu_{B}^{-1}$ determined from our MF
analysis of the magnetic structure. An excellent agreement with
inelastic measurements is apparent.

One possible explanation for the fully  localized nature of the
low-energy spin waves is a strong anisotropy in the effective exchange
interaction between Ni$^{2+}$ and Nd$^{3+}$. We suggest that this
interaction is of Ising type, i.e., can be written as
$J_{{\rm eff}}S^{z}_{{\rm Ni}} S^{z}_{{\rm Nd, eff}}$, ${\bf S}_{{\rm
Nd, eff}}$ being the  effective Nd-spin. If one introduces this type of
coupling between singlet ground state Ni-chains and isolated
paramagnetic rare earth ions at the RPA level, one obtains a total
decoupling of the Ni- and Nd-  transverse degrees of freedom. The Ni-chain  modes
are those of a Haldane system in a {\it static} staggered field. Those
centered on Nd correspond to flipping a single paramagnetic moment,
again in a {\it static} effective field. Interestingly, the low-energy  excitations in Pr$_{2}$BaNiO$_5$ are
qualitatively different: these are acoustic spin waves with a pronounced
dispersion and structure-factor.\cite{Zheludev96PBANO} These
excitations, as well as the higher-energy Haldane-gap modes, are expected
to involve correlated fluctuations of both the Ni and the rare earth
moments. The difference in behavior is, of course, due to the different
electronic configuration of the rare earth ions in Pr$_{2}$BaNiO$_5$ and
Nd$_{2}$BaNiO$_5$ (Pr is not a Kramers ion and in Pr$_{2}$BaNiO$_5$ is, in fact,
an induced-moment system), and the consequent difference in $R$-Ni magnetic
interactions.

Unfortunately, within the MF model one can not make any predictions
regarding the strongly $T$-dependent damping of the 4~meV mode. Without
going into further speculations, we would only like to point out that
the measured magnitude and temperature dependence of $\gamma$ is very
similar to that of the temperature-induced broadening of Haldane-gap
excitations, as measured in Y$_{2}$BaNiO$_{5}$.\cite{Sakaguchi96}  One
possibility is that these quantities may actually be directly related
and represent the same relaxation process.

\subsection{Haldane gap excitations in the Ni chains}

\subsubsection{Polarization}
One set of results that we hoped to obtain from the inelastic thermal
neutron scattering experiments are separate temperature dependencies of
the Haldane gap energies in each of the three components of the Haldane
triplet. Of particular interest is the $T$-dependence of the gap in the
longitudinal mode, i.e., the one polarized along the ordered moments on
the Ni-sites. It has been predicted that the rate of increase of the
longitudinal spin gap is three times larger than that of the two
transverse gaps.\cite{MZ}. As mentioned above, experimental difficulties
make   a complete polarization analysis impossible. At present, the only
available  results relevant to the multiplicity and polarization of
Ni-chain gap excitations in \nd\ is the temperature dependence of
integrated intensities  measured at ${\bf Q}=(1.5,0,0)$ and ${\bf
Q}=(0.5,0,4.2)$ (Fig.~\ref{vst}). To within experimental error these
intensities are the same at all temperatures. At no instance do we see
any distinct splitting in the energy gap. If we assume that the modes
remain polarized along the principal crystallographic directions, as
dictated by the local symmetry of the Ni-sites,  we are forced to
conclude that the three  components of the triplet evolve with
temperature in exactly the same way.

The assumption however is expected to be valid only at $T\gtrsim T_{N}$.
As the magnetic order-induced shift in the Haldane gap energies becomes
larger than the initial anisotropy-induced splitting of the triplet
(roughly 1~meV, according to Refs.~\onlinecite{Xu96,Sakaguchi96}), the
proper choice of polarization axes is governed by the direction of the
Ni$^{2+}$ moments, rather than the crystallographic symmetry directions.
In the magnetically ordered state the three excitations are polarized
along the $x'$, $y'$, and $z'$ axes, where $y'$ runs along the $[010]$
direction, $z'$ is chosen along the direction of ordered moments on the
Ni sites and thus forms an angle of roughly $35^{\circ}$ with the
$c$-axis of the crystal, and $x'$ complete the orthogonal reper,
respectively. The actual situation is even more complex, since two types
of magnetic domains with the Ni$^{2+}$ moments tilted to the left or to
the right of the $c$-axis should be present in a macroscopic sample. As
a result all three modes contribute to inelastic scattering at both
${\bf Q}=(1.5,0,0)$ and ${\bf Q}=(0.5,0,4.2)$, with partial intensities
0.33, 1, 0.67, and 0.63, 1, 0.37, correspondingly.  We have concluded
that our data are not inconsistent with seeing only the two transverse
excitations (polarized along $x'$ and $z'$, respectively). It is
therefore entirely possible that for some reason we fail to see the
longitudinal excitation alltogether.

Definitive conclusions concerning the multiplicity and polarization of
the gap excitations are premature. At this stage however, we lean
towards interpreting our data in terms of seeing only the transverse
Haldane excitations in our scans. First of all, the
longitudinal gap is expected to increase with decreasing $T$ at least
three times as fact as the transverse gaps. Below $T=40$~K it  is
expected to be already outside the range of our inelastic scans. Second,
the observed temperature dependence of the Haldane gap energies is in
striking quantitative agreement with theoretical predictions for
transverse modes (see next section). Third, it appears plausible that
the appearance of an ordered moment on the Ni-sites leads to the
longitudinal mode being overdamped. Hopefully, the use of the very large
sample B in future unpolarized and polarized neutron scattering
experiments will help us resolve this issue.

\subsection{Behavior of the Haldane gap energy}
As discussed in the previous section, our results for the low-energy
spin waves indicate that magnetic
excitations involving the Ni-spins are totally decoupled from those in the Nd-subsytem
and are indistinguishable from transverse Haldane gap excitations that
occur in isolated Haldane chains in the presence of a {\it static}
staggered field. The HC/SF model can thus be used in reference to spin
dynamics, and not only the static magnetic properties of the system.
Based on this assumption (that, at the time, was not backed by
experimental results pertaining to the low-energy spin waves in the
system), the increase of the Haldane gap energy in the magnetically
ordered state in
\ndy\ was qualitatively explained in Ref.~\onlinecite{MZ} In this work
the staggered-field dependence of the Haldane gap energies for a $S=1$
1-D Heisenber antiferromagnet was derived through the use the use of a
Ginsburg-Landau-like description of an isolated Haldane spin chain, the
so-called $\phi^4$-model, first introduced in this context by
Affleck.\cite{Affleck89,Affleck-Wellman92}

In this approach an isolated Haldane spin chain  is  characterized by
the space-dependent vector of local staggered magnetization
$\vec{\phi}(x,t)$. The  Lagrangian of the system is written as a series
expansion in $\vec{\phi}(x)$ and its derivatives:
\begin{equation}
{{\cal L}}=\int dx \left[ \frac{1}{2v}
(\frac{\partial\vec{\phi}}{\partial t})^2
- \frac{v}{2} (\frac{\partial\vec{\phi}}{\partial x})^2- \frac{\Delta^2}{2 v}
\vec{\phi}^2-\lambda |\vec{\phi}|^4+ \mbox{{\rm higher-order terms}}\right]\label{lag}
\end{equation}
The  quadratic terms describe a triplet of degenerate non-interacting
gap excitations (energy gap $\Delta$) with a linear dispersion (spin
wave velocity $v$). A non-zero staggered-field dependence of the
excitation energies appears upon introducing the $\lambda|\vec{\phi}|^4$-term,
that represents pair-repulsion between magnons, and other higher-order
terms. In the presence of a staggered field, the change in gap energies,
defined as $\delta(H_{\pi})\equiv
\case{[\Delta(H_{\pi})]^2-\Delta_0^2}{\Delta_0^2}$, $\Delta_0$ being the gap
energy in the absence of any staggered field, for small fields is
proportional to the square of the induced static staggered magnetic
moment and the strength of magnon repulsion $\lambda$. In application
to $R_2$BaNiO$_5$ systems, $\delta(H_{\pi})$ is
thus expected to scale proportionately to $[m^{({\rm Ni})}]^2$, as long
as the ordered moment is small. This linear relation was confirmed for
all previously studies \ndy\ systems.\cite{MZ,Zheludev97INV}

The data obtained in the present work, particularly   the more
accurately measured temperature dependence of the Ni-sublattice
magnetization, can be used to analyze the relation between $\Delta$ and
$m^{({\rm Ni})}$ in greater detail. In order to do so, we first have to
compensate for the intrinsic weak $T$-dependence of the Haldane gap
energy. This is done by replacing $\Delta_{0}$ by $\Delta_{0}(T)$, the
temperature-dependent gap in the absence of staggered field. For
$\Delta_{0}(T)$ we can take the Haldane gap measured as a function of
temperature in Y$_{2}$BaNiO$_{2}$, that, for practical purposes, we
shall approximate by the following empirical
formula:\cite{Ma92,Zheludev96-NINAZ}
\begin{equation}
\Delta_{0}(T)=\Delta_{0}(0)+\sqrt{\beta T}\exp(-\Delta_{0}(0)/\kappa T)
\end{equation}
The coefficients $\Delta_{0}(0)=9.21$~meV and $\beta=1.0$~meV$^2$/K were
obtained by fitting this form to the experimental data for
Y$_{2}$BaNiO$_{2}$.\cite{Yokoo95,Sakaguchi96} This formula is used to
plot $\delta(H_{\pi})$ as a function of $[m^{({\rm Ni})}]^2$ for all
four
\ndy\ systems studied (Fig.~\ref{maslov}). A perfect data collapse is
obtained. Also apparent is the slow upward curvature of this function
that could not be clearly seen when only powder data were
available.\cite{MZ}

In Ref.~\onlinecite{MZ}  no quantitative predictions regarding the slope
of the linear relation between $\delta(H_{\pi})$ and $[m^{({\rm
Ni})}]^2$, i.e., the coefficient $\lambda$ in Eq.~\ref{lag}, were made.
Instead, $\lambda$ was treated as an adjustable parameter. In a more
refined approach $\lambda$ is to be determined independently from the
properties of the Renormalization Group fixed point. This coefficient
was recently calculated numerically.\cite{Pelissetto97} The
non-linearity in the relation between $\delta(H_{\pi})$ and $[m^{({\rm
Ni})}]^2$ can be accounted  by including higher-order terms into the
$\phi$-expansion of the Lagrangian. The corresponding coefficients, up
to the 8th power in $\vec {phi}$, were also recently
computed.\cite{Pelissetto98} These numerical results were then utilized
to calculate the staggered-moment dependence of the gap energy for both
transverse-polarized and longitudinal Haldane excitations:\cite{newMZ}
\begin{eqnarray}
\frac{\Delta_{\bot}^2-\Delta_{0}^2}{\Delta_{0}^2}=0.395 (m^{({\rm
Ni})})^2+0.156 (m^{({\rm Ni})})^4+0.049(m^{({\rm Ni})})^6\label{tran}\\
\frac{\Delta_{\|}^2-\Delta_{0}^2}{\Delta_{0}^2}=1.185 (m^{({\rm
Ni})})^2+0.78 (m^{({\rm Ni})})^4+0.343(m^{({\rm Ni})})^6\label{long}
\end{eqnarray}
These equations have {\it no adjustable parameters} and for \nd\ systems
yield the curves shown in solid lines in Fig.~\ref{maslov}. We see that
our data almost perfectly agrees  with the prediction for the transverse
spin gap. The deviations at large $m^{({\rm Ni})}$ are expected, as the
series in Eqs.~(\ref{tran},\ref{long}) are terminated at the 3rd term.
Assuming that what we observe in our experiments are transverse gap
excitations, the HC/SF model thus gives {\it quantitatively correct}
predictions for the temperature dependence of the gap energies.

\section{conclusion}
The results of three independent series of experiments, powder neutron
diffraction, and thermal- and cold- neutron inelastic scattering, are
all consistent with our HC/SF model for $R_2$BaNiO$_5$ compounds at the
quantitative level. The interactions between the Ni and rare earth
subsystems in \ndy\ can be to a very good approximation treated
at the MF level. This description appears to  apply very well not only
to static magnetic properties, but also to the spin dynamics. The
dispersionless single-ion character of low-energy spin waves is a
manifestation  of an effective separation of the Ni-chain spin dynamics
from that of the Nd-subsystem.

\acknowledgements
The authors would like to thank P. Convert, B. Malaman, J. L. Soubeyroux
and D. Neuman for their help with neutron scattering experiments.  This
study was in part supported by the U.S.-Japan Cooperative Program on
Neutron Scattering, a Grant-in-Aid for Scientific Research from the
Ministry of Education, Science and Culture Japan and The Science
Research Fund of Japan Private School Promotion Foundation.  Work at
Brookhaven National Laboratory was carried out under Contract No. 
DE-AC02-76CH00016, Division of Material Science, U.S.\ Department of
Energy. Studies at NIST were partially supported by the NSF under
contract No. DMR-9413101.


\begin{figure}
\caption{Typical measured (symbols) and calculated (solid line) powder diffraction profiles for
in \protect\ndy\ samples with $x=1$ (a), $x=0.75$ (b), $x=0.5$ (c), and
$x=0.25$ (d), and the corresponding residuals (obs - calc).  For $x=1$, the
full diffraction pattern (nuclear and magnetic) is shown.  For the three other compounds,
the non-magnetic scattering profile was measured just above the N\'{e}el
temperatures and subtracted from shown scans.  The ticks indicate positions of
Bragg reflections,  both nuclear (N) (in the $x=1$ compound) and magnetic
(M).  In the magnetically ordered phase of the $x=0.25$
compound (d), additional intensity is also observed at nuclear Bragg peak
positions (marked with an $\ast$).}
\label{difdata}
\end{figure}

\begin{figure}
\caption{ A schematic view of the magnetic structure of Nd$_{2}$BaNiO$_{5}$.
A single crystallographic unit cell is shown. The magnetic moments of
both Ni$^{2+}$ and Nd$^{3+}$ are confined to the $(a,c)$
crystallographic plane. The magnetic propagation vector is $[\case{1}{2}
0 \case{1}{2}]$.}
\label{cartoon}
\end{figure}

\begin{figure}
\caption{Measured ordered moment on the Ni (solid circles) and Nd (open circles) sites
in \protect\ndy\ samples for $x=1$, $x=0.75$, $x=0.5$, and $x=0.25$,
plotted against temperature (symbols). Solid lines are calculated from
the mean field model described in the text.}
\label{MvsT}
\end{figure}

\begin{figure}
\caption{Measured angle formed by the  Ni (solid circles) and Nd (open circles)
ordered moments with the $c$ crystallographic axis in \protect\ndy\
samples for $x=1$, $x=0.75$, $x=0.5$, and $x=0.25$, plotted against
temperature. Solid lines are guides to the eye.}
\label{PvsT}
\end{figure}

\begin{figure}
\caption{Sample inelastic scans measured in single-crystal Nd$_2$BaNiO$_5$ (sample A)
at the 1-D antiferromagnetic zone-centers (open symbols) and background
scans taken slightly off these positions (solid symbols) at $T=50$~K.
The shaded areas indicates the scan range contaminated by a $2k_i-3k_f$
spurion.}
\label{exdata}
\end{figure}

\begin{figure}
\caption{Sample inelastic scans (background subtracted) measured in single-crystal
Nd$_2$BaNiO$_5$ (sample A) at the 1-D antiferromagnetic zone-centers
(open symbols). The solid lines are fits with a model cross section as
described in the text.}
\label{exsub}
\end{figure}

\begin{figure}
\caption{Sample inelastic scans (background subtracted) measured in single-crystal
Nd$_2$BaNiO$_5$ (sample B) at the 1-D antiferromagnetic zone-centers
(open symbols) at several temperatures. The solid lines are fits with a
model cross section as described in the text.}
\label{exsub2}
\end{figure}

\begin{figure}
\caption{Measured temperature dependencies of the gap energy (top) and
energy-integrated intensity (bottom) of Ni-chain Haldane excitations in
single-crystal Nd$_2$BaNiO$_5$. Triangles and circles show data obtained
in the two samples studied. The solid lines are guides to the eye.}
\label{vst}
\end{figure}

\begin{figure}
\caption{Sample inelastic scans measured in Nd$_2$BaNiO$_5$ (sample B) at the
magnetic Bragg position (0.5,0,1.5) using thermal (top) and cold
(bottom) neutrons at $T=10$~K. Apart from the resolution-limited feature
at 4 meV energy transfer, no inelastic scattering is observed above
background level up to 8 meV energy transfer. Solid lines are guides to
the eye.}
\label{nowave}
\end{figure}

\begin{figure}
\caption{Sample inelastic scans measured in Nd$_2$BaNiO$_5$ (sample B) at the
magnetic Bragg  zone-center (a) and zone-boundaries (b,c) illustrate the
absence of any dispersion in the 4 meV excitation at $T=10$~K. Solid
lines are Gaussian fits.}
\label{spinsdata1}
\end{figure}

\begin{figure}
\caption{Temperature evolution of the 4 mev excitation measured in Nd$_2$BaNiO$_5$ (sample B)
at the magnetic Bragg position (0.5,0,1.5). The solid lines represent
empirical fits as described in the text. At $T>T_{N}=48$~K the
excitation is totally unobservable.}
\label{spinsdata2}
\end{figure}

\begin{figure}
\caption{Measured temperature dependence of the energy squared (open circles) and full energy
width at half maximum (solid circles) of the 4 meV excitation in
Nd$_2$BaNiO$_5$ (sample B). The solid line is a theoretical curve
described in the text.}
\label{spinsresults}
\end{figure}

\begin{figure}
\caption{Measured ordered Ni$^{2+}$-moment in \protect\ndy\ compounds plotted
against the average ordered moment on the $R$-sublattice (symbols, bottom
axis). This  dependence is interpreted as the staggered magnetization
curve for an isolated Haldane spin chain. The estimated staggered
exchange field acting on the Ni-subsystem is shown on the top axis. The
solid line is an empirical fit to the experimental points, as described
in the text.}
\label{univ}
\end{figure}

\begin{figure}
\caption{Measured increase in the Haldane gap energy in \protect\ndy\ compounds relative
to that in Y$_2$BaNiO$_5$ plotted as a function of the square of the
ordered staggered moment on the Ni-chains (symbols). For  \nd\
single-crystal data obtained in this work is shown. For the three other
materials the data are taken  from Ref.~\protect\onlinecite{Yokoo97NDY}.
The solid lines are theoretical predictions for the longitudinal and
transverse gaps (Ref.~\protect\onlinecite{newMZ}).}
\label{maslov}
\end{figure}

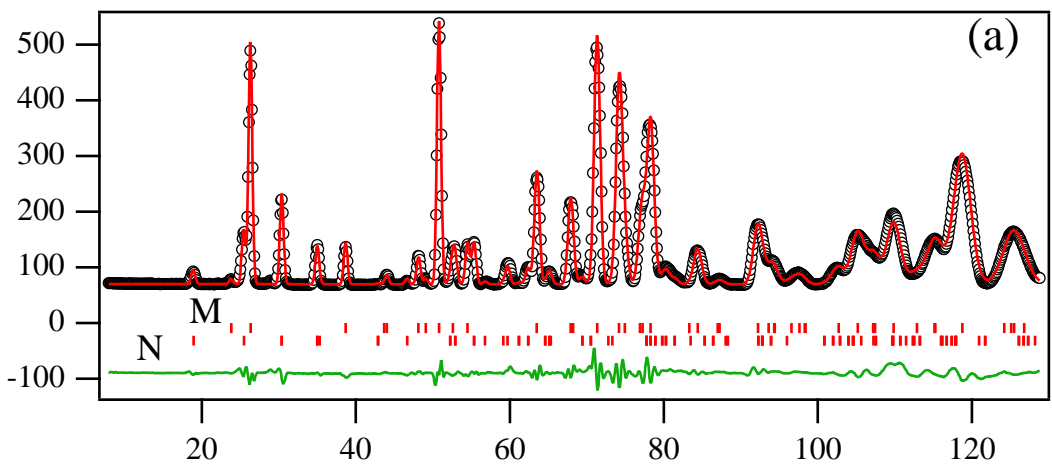
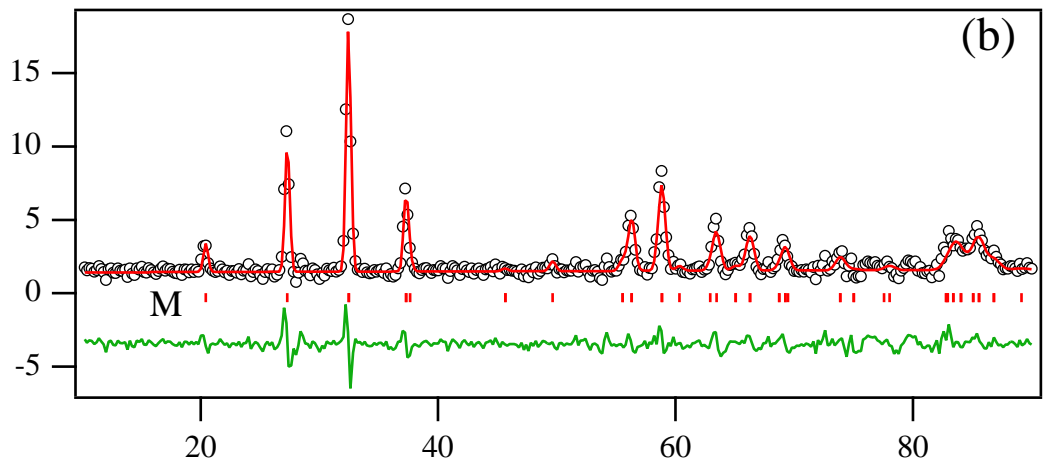
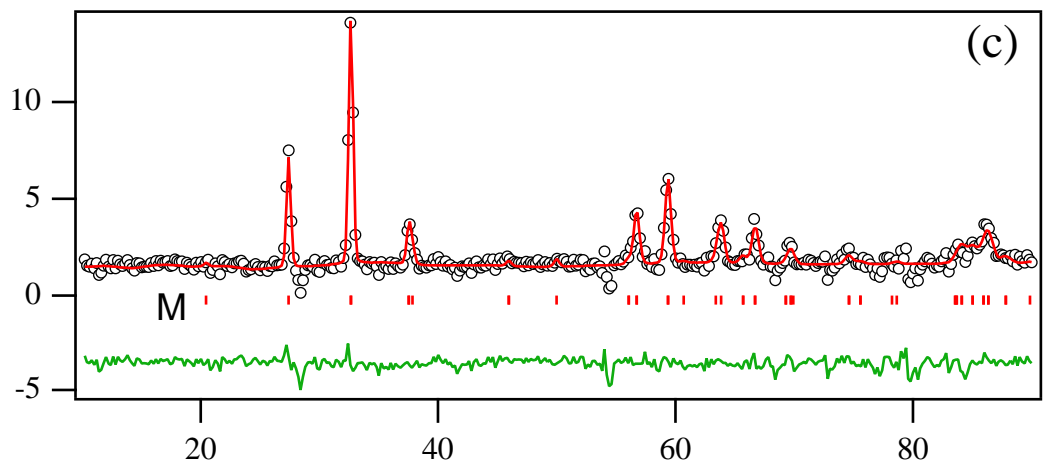
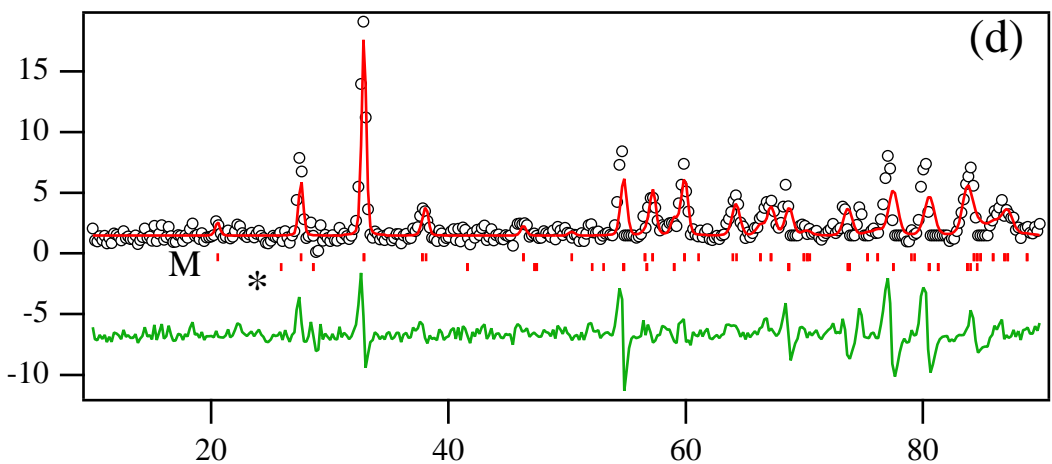
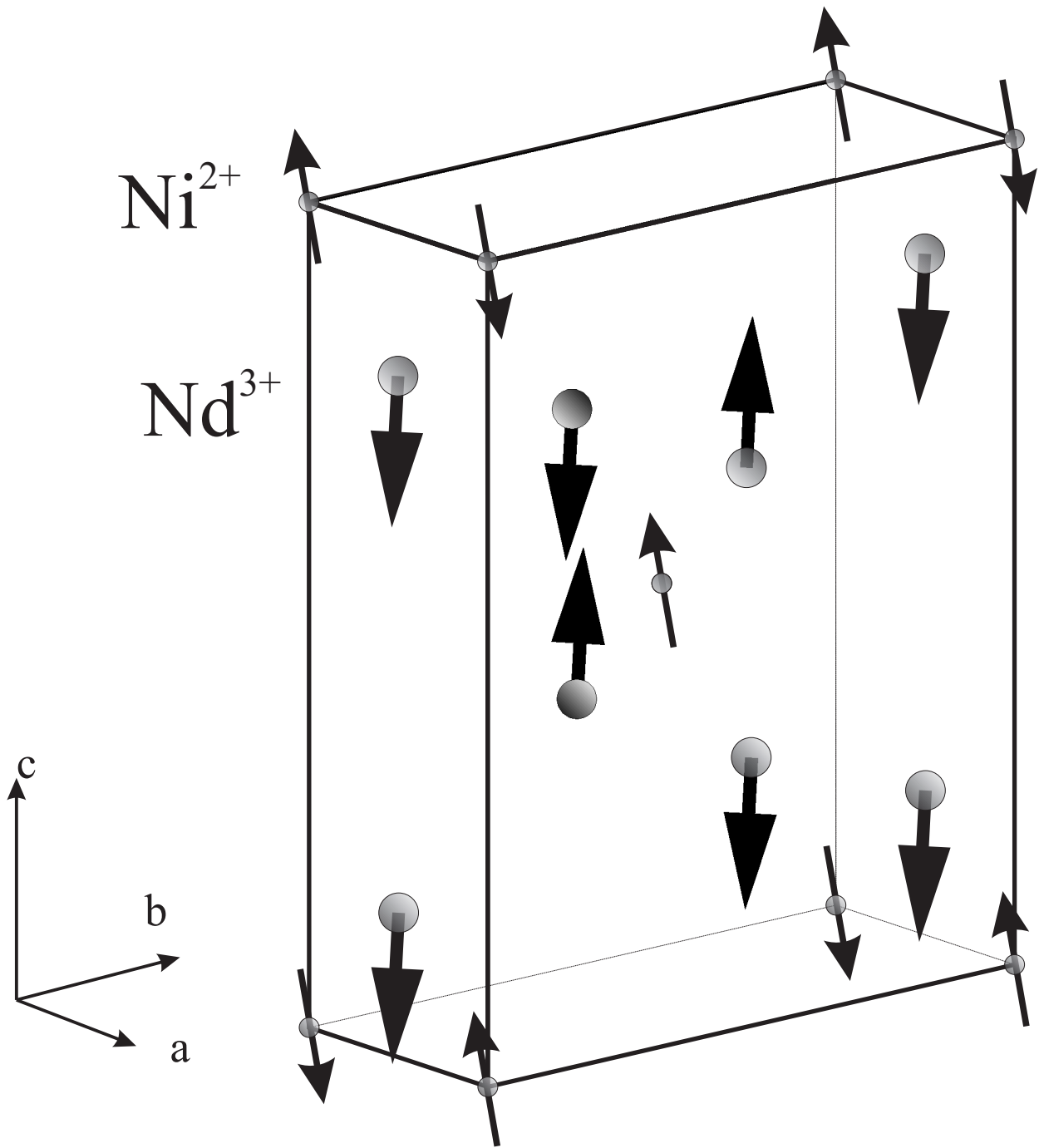
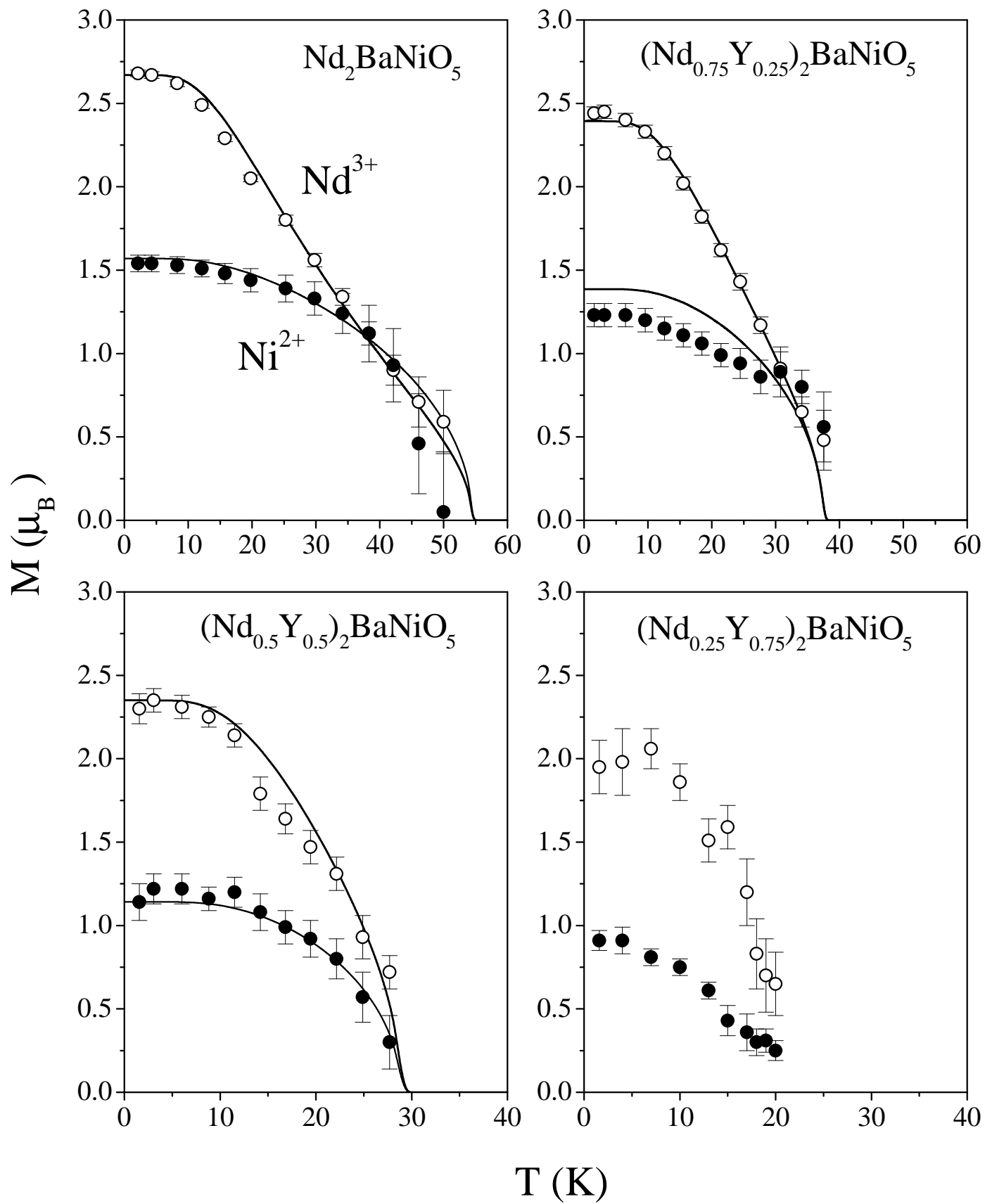
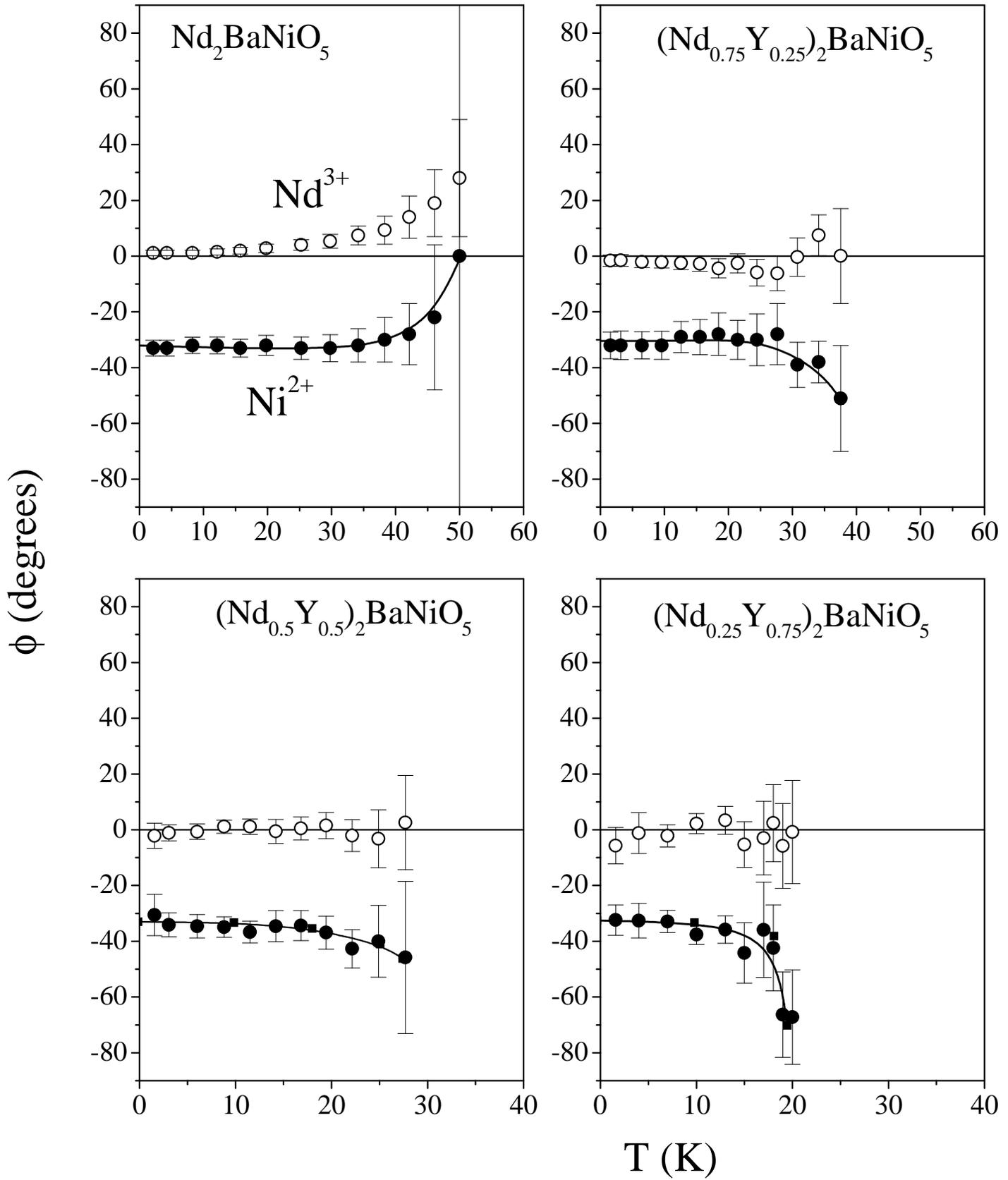
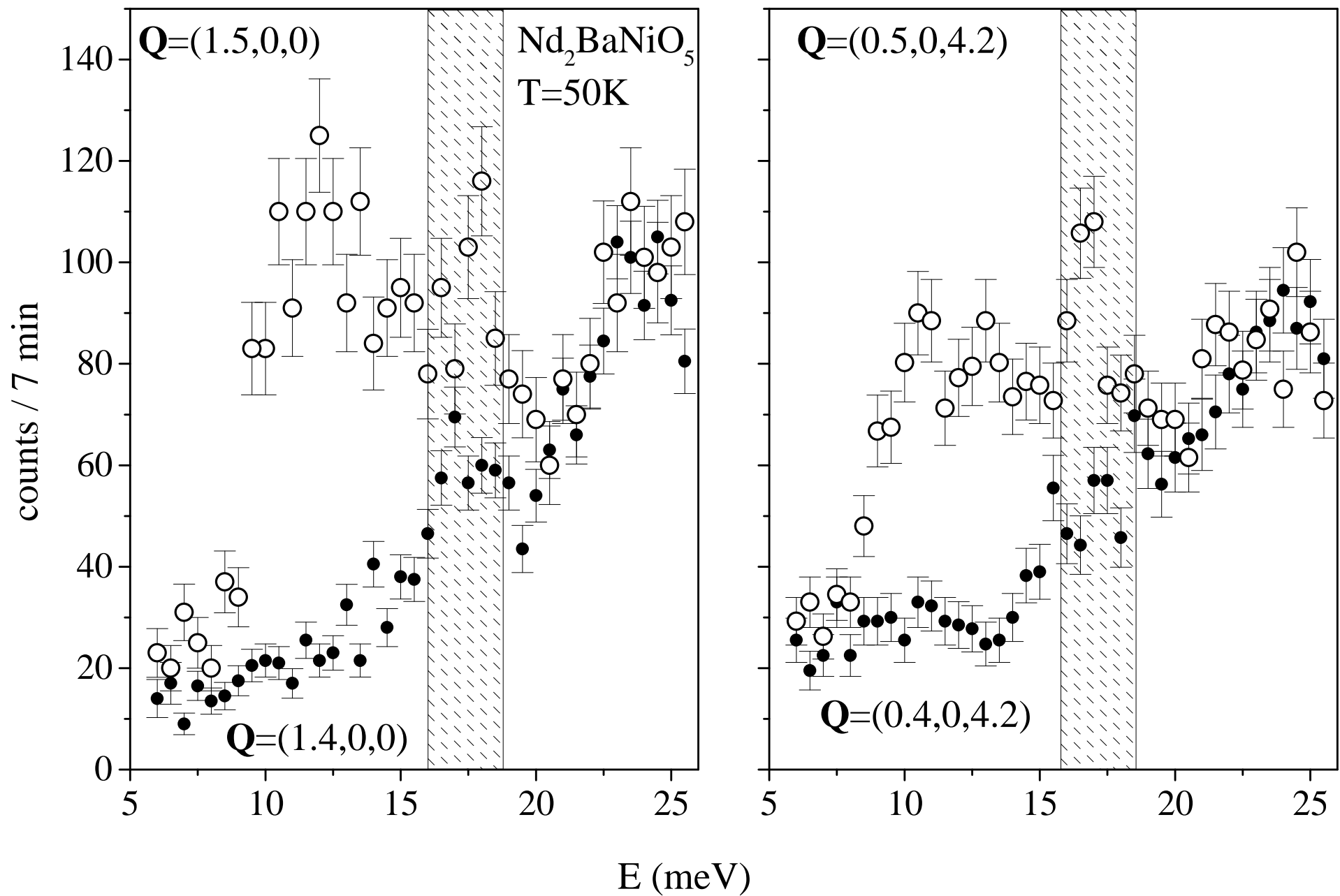
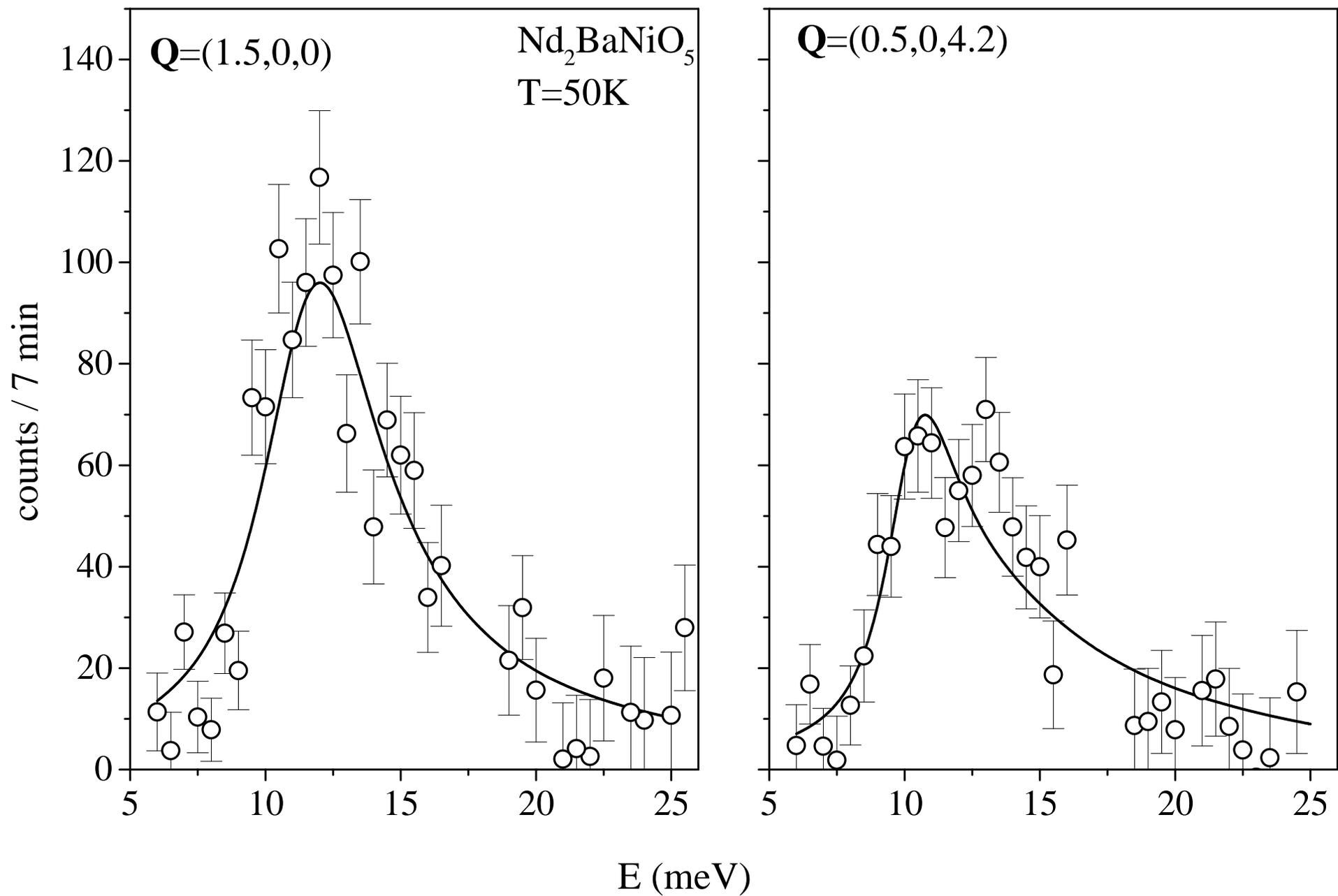
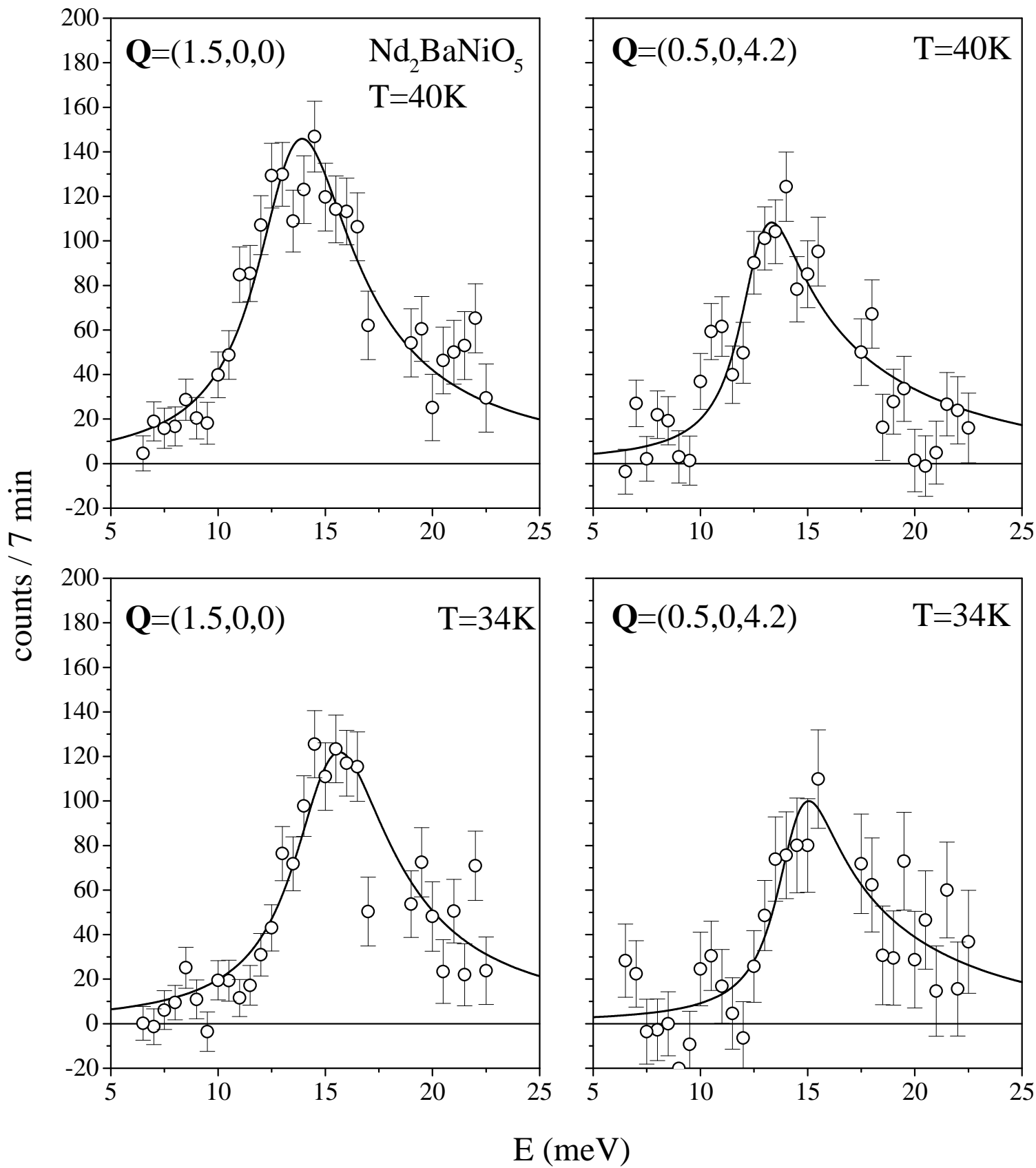
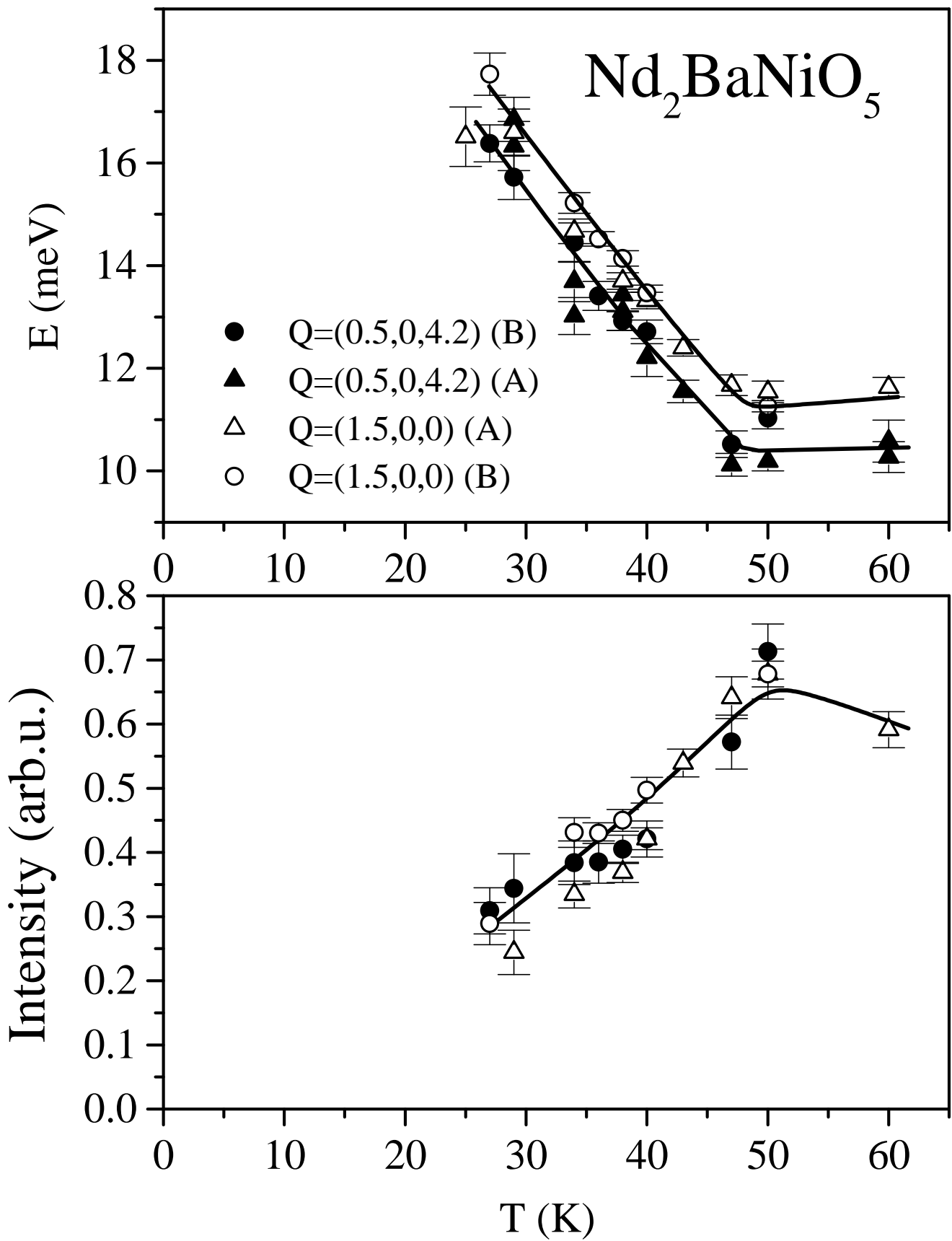
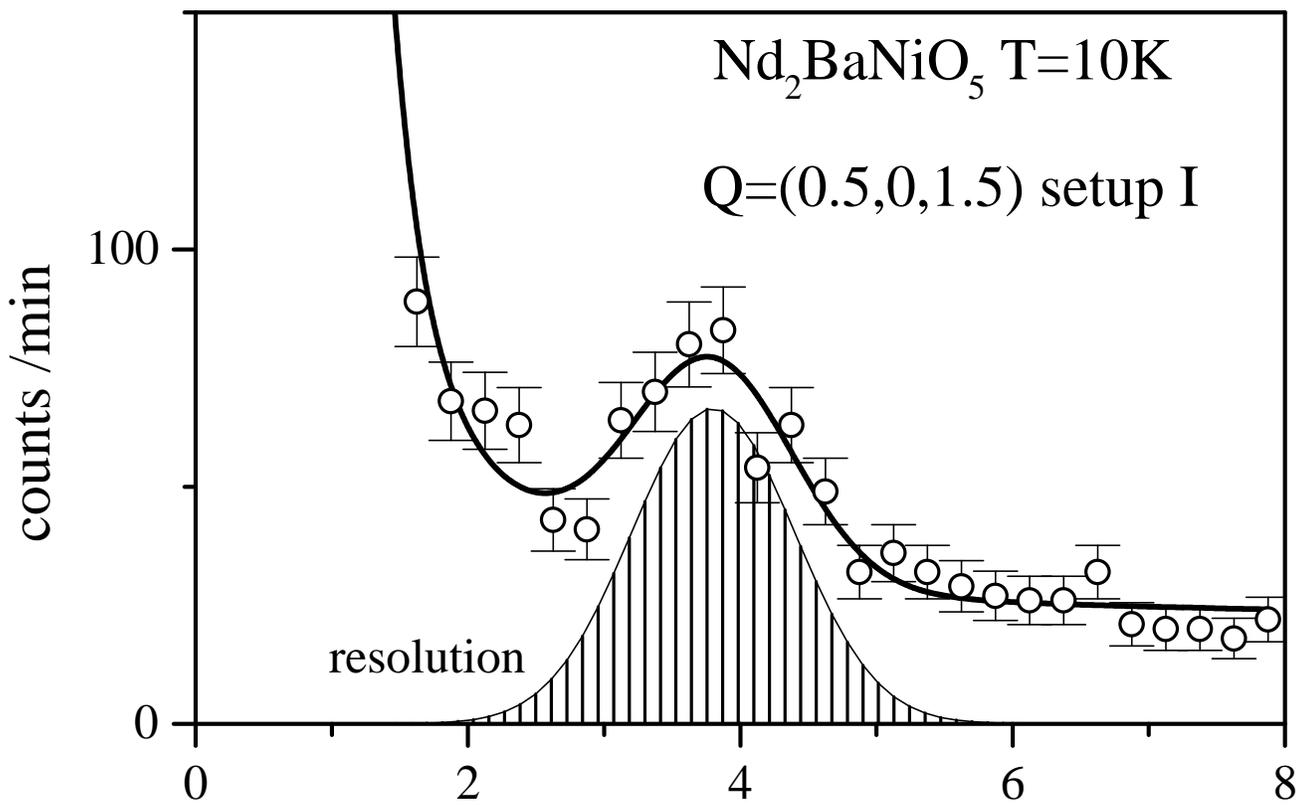
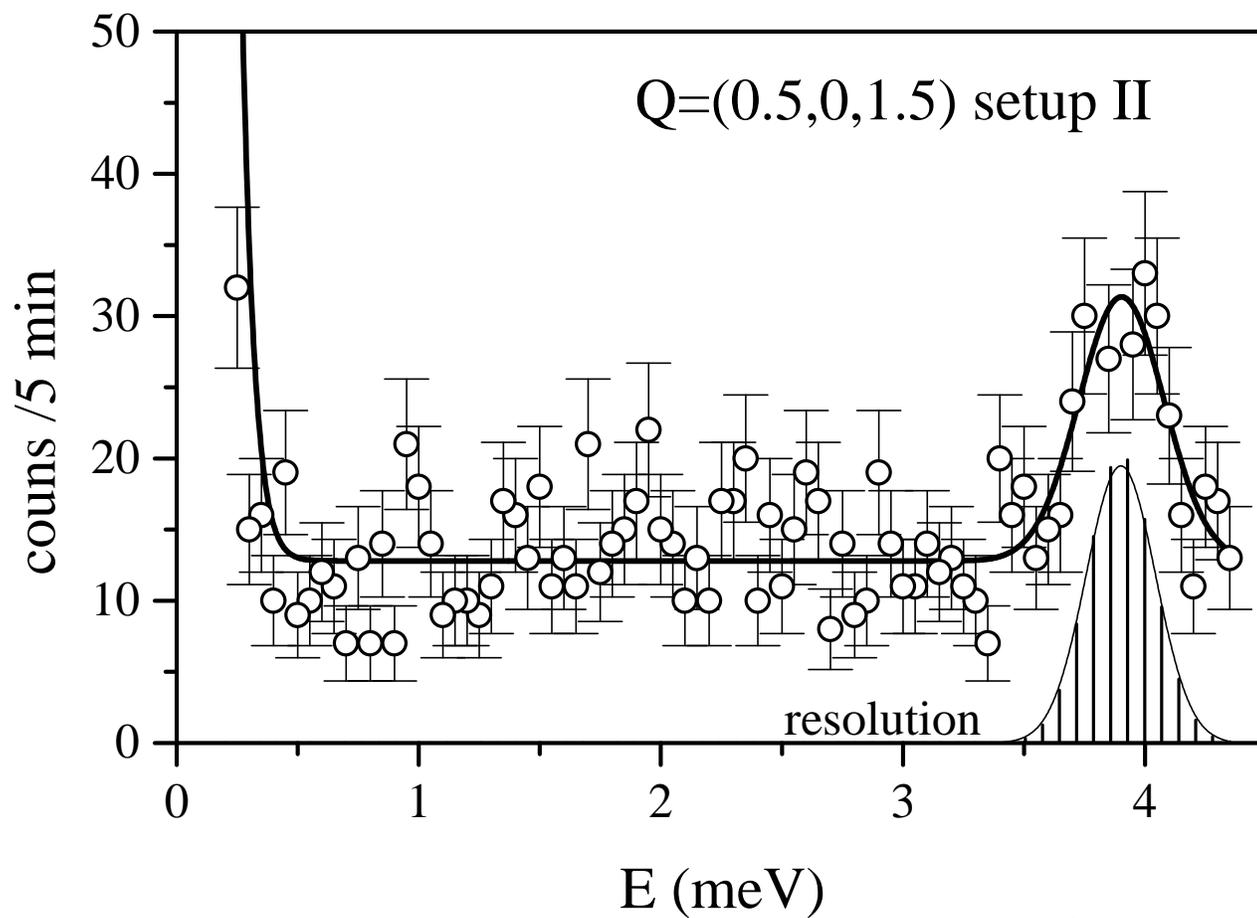
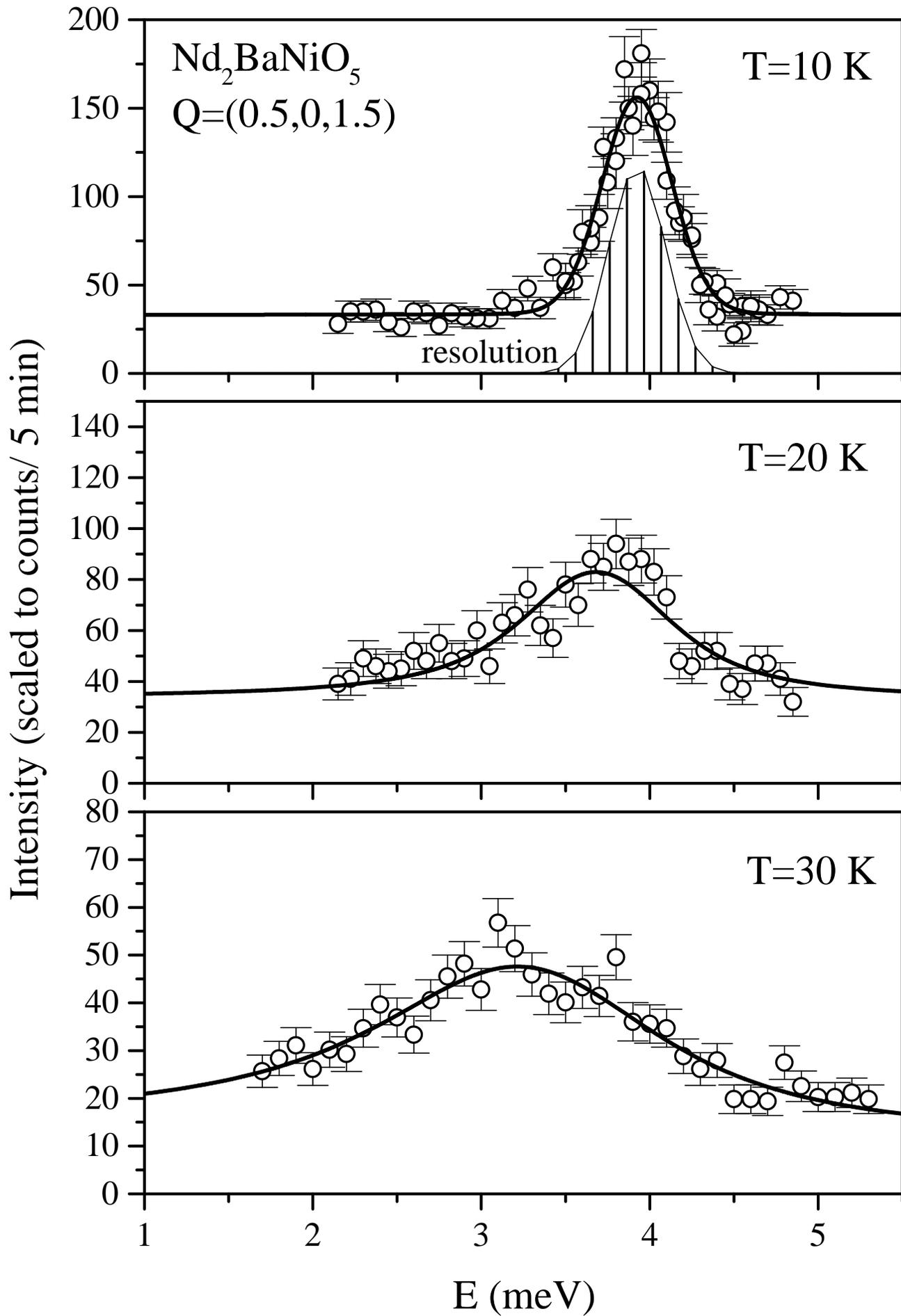
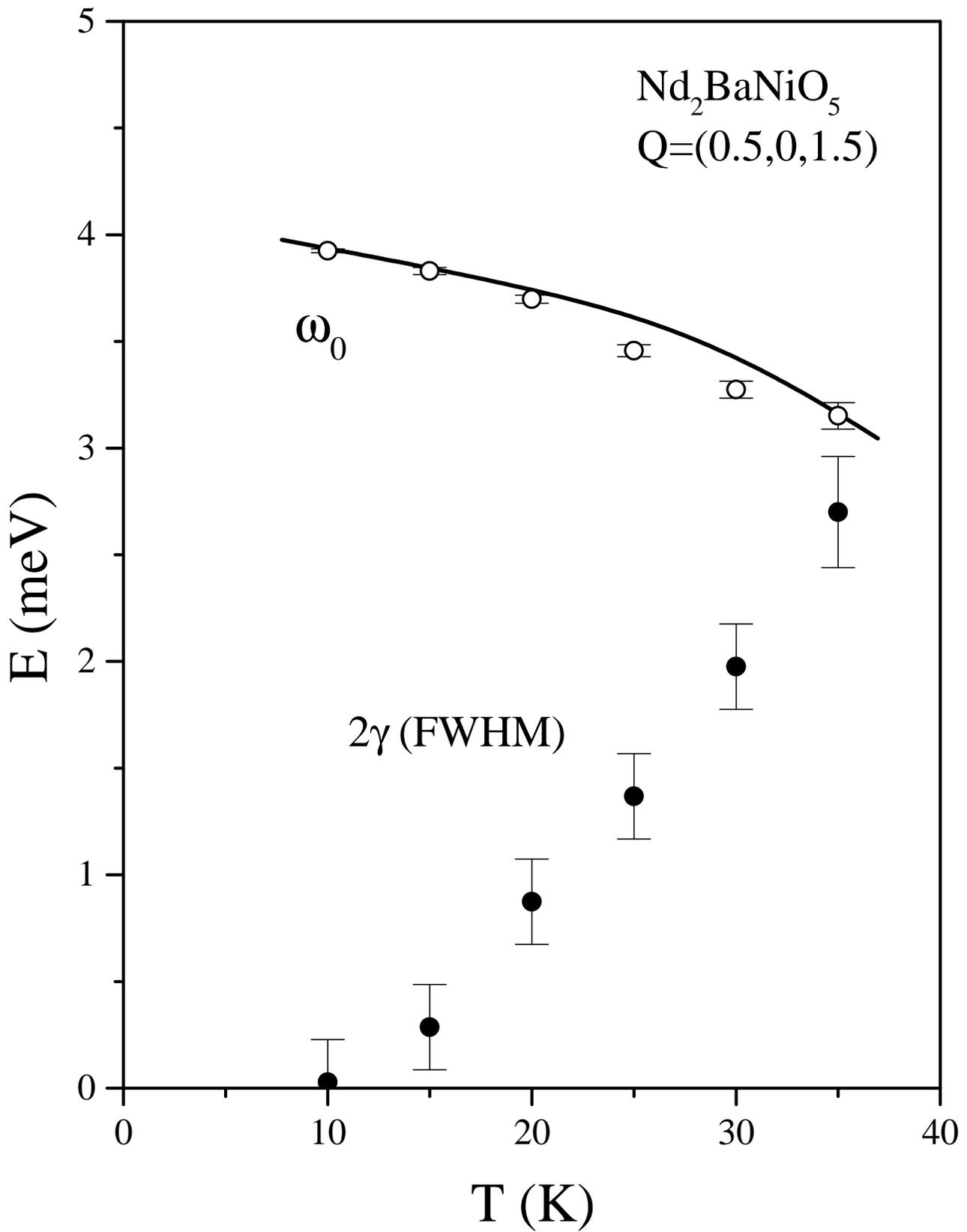
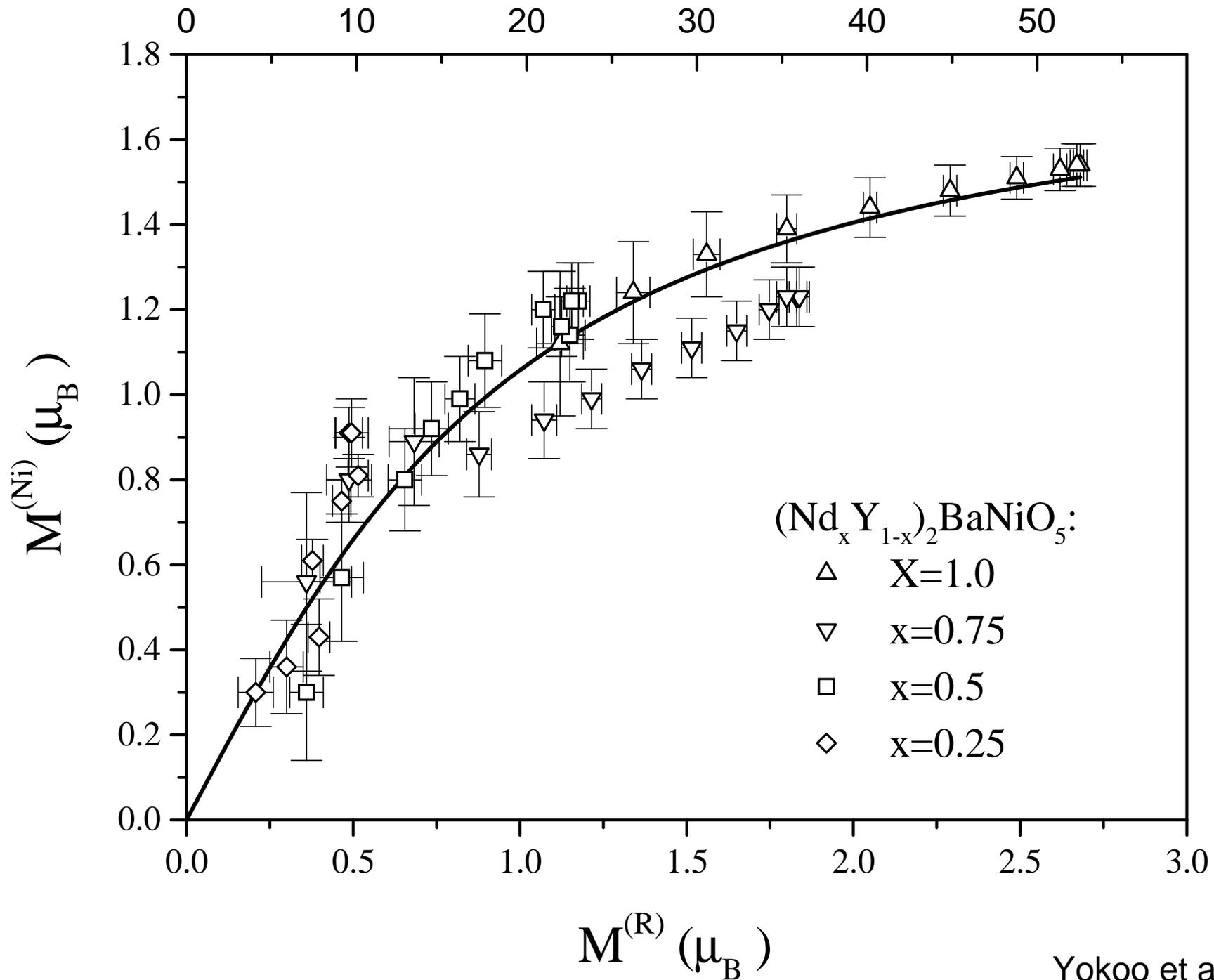
\end{document}